\documentclass[useAMS,usenatbib]{mn2e}

\usepackage{times}
\usepackage{graphicx}
\usepackage{psfig}
\usepackage{epsfig}
\usepackage{amsfonts}
\usepackage{amsmath}
\usepackage{amssymb}
\usepackage{amscd}
\usepackage{multicol}
\usepackage[T1]{fontenc}
\usepackage{aecompl}

\def \anu {1}
\def \caastro {2}
\def \sifa {3}
\def \swinburne {4}
\def \aao {5}
\def \melb {6}
\def \unc {7}
\def \uq {8}
\def \uwa {9}
\def \durham {10}
\def \uha {11}
\def \monash {12}
\def \eso {13}
\def \macquarie {14}
\def \aip {15} 

\title[Generating spectroscopic datacubes]{The SAMI Galaxy Survey: Cubism and covariance, putting round pegs into square holes}
\author[The SAMI Galaxy Survey team]
{\parbox{\textwidth}{ \raggedright R.\,Sharp,$^{\anu,\caastro}$\thanks{E-mail: Rob.Sharp@anu.edu.au}
J.\,T.~Allen$^{\sifa,\caastro}$,
L.\,M.\,R.~Fogarty$^{\sifa,\caastro}$,
S.\,M.~Croom$^{\sifa,\caastro}$, 
L.~Cortese$^{\swinburne}$, 
A.\,W.~Green$^{\aao}$,
J.~Nielsen$^{\anu}$,
S.\,N.~Richards$^{\sifa,\caastro,\aao}$,
N.~Scott$^{\sifa,\caastro}$,
E.\,N.~Taylor$^{\melb}$,
L.\,A.~Barnes$^{\sifa}$, 
A.\,E.~Bauer$^{\aao}$,
M.\,~Birchall$^{\aao}$,
J.~Bland-Hawthorn$^{\sifa,\caastro}$,
J.\,V.~Bloom$^{\sifa,\caastro}$,
S.~Brough$^{\aao}$,
J.\,J.,~Bryant$^{\sifa,\caastro,\aao}$,
G.\,N.~Cecil$^{\unc}$,
M.~Colless$^{\anu}$,
W.\,J.~Couch$^{\aao}$,
M.\,J.~Drinkwater$^{\uq}$,
S.~Driver$^{\uwa}$,
C.~Foster$^{\aao}$,
M.~Goodwin$^{\aao}$,
M.~L.~P., Gunawardhana$^{\durham}$,
I.-T.~Ho$^{\uha}$,
E.\,J.~Hampton$^{\anu}$,
A.\,M.~Hopkins$^{\aao}$,
H.~Jones$^{\monash}$,
I.\,S.~Konstantopoulos$^{\aao,\caastro}$,
J.\,S.,~Lawrence$^{\aao}$,
S.\,K.~Leslie$^{\anu}$,
G.\,F.~Lewis$^{\sifa}$,
J.~Liske$^{\eso}$,
\'A.R.~L\'opez-S\'anchez$^{\aao,\macquarie}$
N.\,P.\,F.~Lorente$^{\aao}$,
R.~McElroy$^{\sifa,\caastro}$
A.\,M.~Medling$^{\anu}$,
S.~Mahajan$^{\uq}$,
J.~Mould$^{\swinburne}$,
Q.~Parker$^{\aao,\macquarie}$,
M.\,B.~Pracy$^{\sifa}$,
D.~Obreschkow,$^{\uwa}$,
M.\,S.~Owers$^{\aao}$,
A.\,L.~Schaefer$^{\sifa,\caastro,\aao}$,
S.\,M~Sweet$^{\anu,\uq}$,
A.\,D.~Thomas$^{\uq}$,
C.~Tonini$^{\melb}$,
C.\,J.~Walcher$^{\aip}$,
}\vspace{0.4cm}\\
\parbox{\textwidth}{$^{\anu}$ Research School of Astronomy \& Astrophysics, Australian National University, Canberra, ACT 2611, Australia\\
$^{\caastro}$ ARC Centre of Excellence for All-sky Astrophysics (CAASTRO)\\
$^{\sifa}$ Sydney Institute for Astronomy (SIfA), School of Physics, The University of Sydney, NSW 2006, Australia\\
$^{4}$ Centre for Astrophysics \& Supercomputing, Swinburne University of Technology, Mail H30 P.O. Box 218, Hawthorn, VIC 3122, Australia\\
$^{\aao}$ The Australian Astronomical Observatory, PO Box 915, North Ryde, NSW, 1670, Australia\\
$^{\melb}$ School of Physics, The University of Melbourne, VIC 3010, Australia\\
$^{\unc}$ Dept. Physics \& Astronomy, University of North Carolina, Chapel Hill, NC 27599, USA\\
$^{\uq}$ School of Mathematics and Physics, University of Queensland, Brisbane, QLD 4072, Australia\\
$^{\uwa}$ ICRAR M468, UWA, 35 Stirling Highway, Crawley, WA 6009, Australia\\
$^{\durham}$ Institute for Computational Cosmology, Department of Physics, Durham University, South Road, Durham DH1 3LE, UK\\
$^{\uha}$ Institute for Astronomy, University of Hawaii, 2680 Woodlawn Drive, Honolulu, HI 96822, USA\\
$^{\monash}$ School of Physics, Monash University, VIC 3800, Australia\\
$^{\eso}$ European Southern Observatory, Karl-Schwarzschild-Str. 2, 85748 Garching bei M\"unchen, Germany\\
$^{\macquarie}$ Department of Physics and Astronomy, Macquarie University, NSW 2109, Australia\\
$^{\aip}$ Leibniz-Institut f\"ur Astrophysik Potsdam (AIP), An der Sternwarte 16, D-14482 Potsdam, Germany
}
}
\begin{document}

\date{Accepted YYYY month DD. Received YYYY month DD; in original form YYYY month DD}

\pagerange{\pageref{firstpage}--\pageref{lastpage}} \pubyear{YYYY}

\maketitle

\label{firstpage}

\begin{abstract}
We present a methodology for the regularisation and combination of sparse sampled and irregularly gridded observations from fibre-optic multi-object integral-field spectroscopy. The approach minimises interpolation and retains image resolution on combining sub-pixel dithered data. We discuss the methodology in the context of the Sydney-AAO Multi-object Integral-field spectrograph (SAMI) Galaxy Survey underway at the Anglo-Australian Telescope. The SAMI instrument uses 13 fibre bundles to perform high-multiplex integral-field spectroscopy across a one degree diameter field of view. The SAMI Galaxy Survey is targeting $\sim$3000 galaxies drawn from the full range of galaxy environments. We demonstrate the subcritical sampling of the seeing and incomplete fill factor for the integral-field bundles results in only a 10\% degradation in the final image resolution recovered. We also implement a new methodology for tracking covariance between elements of the resulting datacubes which retains 90\% of the covariance information while incurring only a modest increase in the survey data volume.
\end{abstract}


\section{Introduction}

For two decades, single-fibre multi-object spectrographs dominated galaxy redshift surveys
\citep{huchra99,york00,colless01,eisenstein05,jones09,driver11,drinkwater12}.
More than two million galaxies now have accurate redshift measurments. These surveys have taught us a great deal about large-scale structure and how the bolometric properties of galaxies evolve with cosmic time.
In recent years, broadband photometric surveys have revealed there is much to be learnt from the spatially resolved properties of large galaxy samples \citep{driver06,abazajian09} in particular, how these properties vary with the large-scale environment \citep{blanton09,welikala08}. What is missing from these surveys is knowledge of the kinematics, ages and metallicity of prominent stellar populations across each galalxy, plus the emission line intensities and kinematics that provide insight on the star formation, dynamics and chemical state of the galaxy.

Integral field and Fabry-Perot spectrographs provide the necessary 3D (2D spatial, 1D spectral) information but these are typically designed as single-target instruments
\citep[e.g., HIFI \& PMAS - ][]{bland89,roth05}
and modest survey samples
\citep{veilleux03,sharp10c, rich12, brough13}.
Recent integral field surveys have managed to observe of order 200$-$300 galaxies over a long observing campaign (e.g., SAURON - \citet{bacon01}; ATLAS3D - \citet{cappellari11}; CALIFA - \citet{sanchez12}).

With a view to obtaining integral field data on thousands of galaxies, the SAMI concept was born
\citep{croom12}. This multi-object instrument uses a new kind of fibre bundle
-- the hexabundle \citep{bland-hawthorn11,bryant11} --
in order to achieve spatially resolved 3D spectroscopy of up to 13 galaxies at a time over a 1 degree diameter field. The SAMI
Galaxy Survey will observe more than 3000 galaxies in a 3~yr campaign.
This instrument is proposed to be extended to 
50$-$100 bundles in its next incarnation with a view to obtaining data on a far larger sample of objects
\citep{lawrence12}. One challenge of the hexabundle technique is its irregular fibre format of close-packed circular fibres (Figure~\ref{layout}) which for many applications must be reformatted to give well-formed data for uniform analysis. Indeed the same issues are faced by any imaging system if none integer spaxel shifts are employed in dithering. In this paper we present the resampling solution adopted, along with other processing steps required to remove the instrumental signature, for the SAMI Galaxy Survey.

The code to perform the datacube generation described in this paper is available from the Astrophysics Source Code Library as project asci:1407.006\footnote{Astrophysics Source Code Library, asci:1407.006, http://ascl.net/1407.006 \citep{allen14a}.}.

\section{The Sydney-AAO Multi-object Integral-field spectrograph (SAMI)}
The motivation for the SAMI instrument is described by \citet{croom12} with technical specification provided in \citet{bryant12} and updated in \citet{bryant14b}. The first explorations of its scientific capabilities are provided by \citet{fogarty12,fogarty14,ho14,richards14}.

\begin{figure}
\begin{center}
\includegraphics[width=70mm]{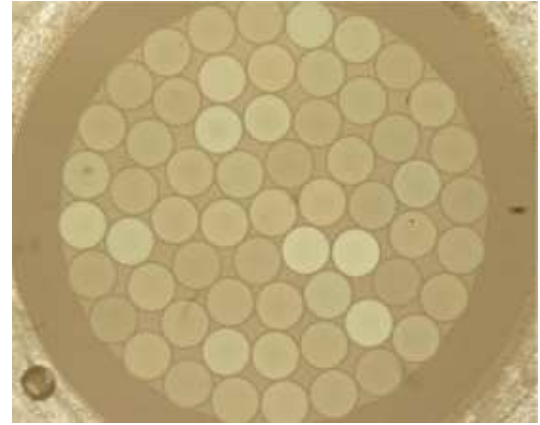}
\includegraphics[width=80mm]{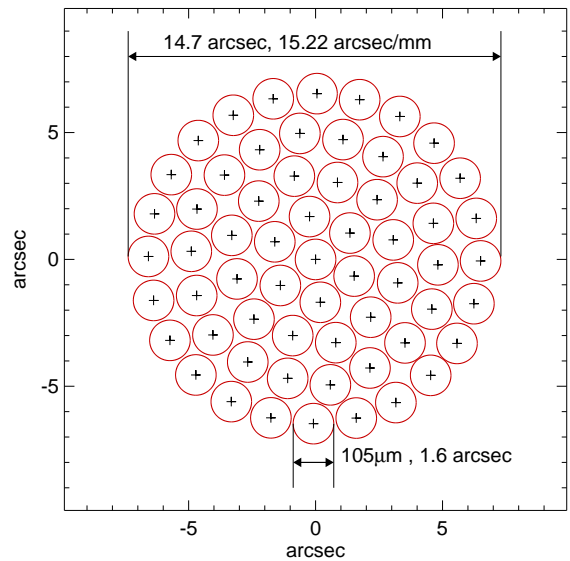}
\caption{\label{layout} The fibre core arrangement is shown for one of the SAMI fibre IFUs. A quality control image of a bundle, taken during production of a hexabundle, is shown along with a graphical fibre mapping
The none uniform fibre illumination pattern in the image is a product of the back-illumination used for the photograph
and is not representative of the final bundle transmission. A number of defocused dust artefacts on the laboratory camera are also seen. Each fibre has a core diameter of 105$\mu$m, with the cladding thinned to 110$\pm$1\,$\mu$m over the first 50\,mm of each fibre.  This allows a tight fibre packing, creating a semi-regular IFU array with a filling factor of 73\%.}
\end{center}
\end{figure}

Briefly, the SAMI system uses lightly fused fibre bundles to create self-contained fibre integral-field units. Each of the 13 SAMI fibre bundles contains a close packed array of 61 optical fibres with individual fibre-core diameters of 1\farcs6 (Figure~\ref{layout}). The confinement of the bundle by a circular outer form arranges the fibres into 4 concentric rings around the central fibre, rather than a hexagonal packing \citep{bryant14a}. The fibre bundles populate the 1$^\circ$ diameter focal plane of the $F/3.4$ triplet corrector at the Anglo-Australian Telescope (AAT) via a plug-plate system. Each fibre-bundle has a fill factor of 73\% over a FoV of $\sim$15\arcsec diameter. The 13 IFU bundles, each of 61 fibres, and 26 independent blank-sky fibres for sky-subtraction, feed the AAOmega spectrograph \citep{saunders04,sharp06}. Using the 580V and 1000R gratings, the dual beam AAOmega spectrograph provides wavelength coverage in two bands, 3700-5700\AA\ at a spectral resolution of R$\sim$1730 and 6250-7350\AA\ at R$\sim$4500. The SAMI galaxy survey will observe more than 3000 galaxies in a 3~year campaign as described in \citet{bryant14b}.

\section{Basic data processing}
A detailed description of the fibre spectroscopy data reduction
software \texttt{2dFdr}, whose usage is common to data taken in all
modes of the AAOmega spectrograph, is presented by \citet{hopkins13}
who provide detail of AAOmega data processing for the
GAMA survey program \citep{driver11}. \texttt{2dFdr} carries out all
the reduction steps up to the point of generating Row-Stacked Spectra
(RSS) which are wavelength calibrated and sky subtracted with the basic
instrumental signatures removed.  The RSS frames are 2D images with 
one row per fibre spectrum (along
with the associated variance information and source details in
image and binary table extensions) for each observation with the SAMI IFUs.  The
description of how these are flux calibrated and converted into
datacubes are given in Sections \ref{fluxcal} and \ref{drizzle}
below. Here we will describe the steps required to produce the RSS
frames within the \texttt{2dFdr} package.  

The first stage is to subtract bias and dark frames to correct a
number of errant CCD pixels.
Both arms of the
dual-beam AAOmega spectrograph suffers from a number of
extended regions of bad columns
whose charge transfer inefficiency effects associated with hot pixels can be
compensated for by bias/dark correction.\footnote{In April 2014, the cosmetically poor blue AAOmega CCD was replaced with new device largely free from such defects. The continued need for full bias/dark correction will be reconsidered as experience is gained with the new array. The red CCD will be upgraded in August 2014.}
An overscan correction is also applied, subtracting the
bias level in each frame.  After this, each frame is divided by a
{\it detector flat} that is generated by averaging (typically $>30$) fibre flats for which the spectrograph has been
defocussed so that the illumination is relatively uniform. These
frames are then filtered to remove large-scale variations, leaving
only smaller-scale pixel-to-pixel flat field variations.  Charge spots due to cosmic rays are 
removed from each individual 
science frame using a tuned implementation of the LaCosmic routine
\citep{vandokkum01,husemann12}. An optimisation of the parameters was
performed to ensure high confidence in cosmic-ray rejection with
minimal impact on fibre spectra. 

The next stage is to trace the fibre locations across the detector
(generating a so-called {\it tramline map} giving the pixel-by-pixel [x,y] location
of each fibre).  This is a crucial step as good extraction of 1D
spectra from the 2D data frame is contingent on accurately mapping the
fibre positions and profiles across the detector.  This is performed
using a fibre flat field frame taken using a quartz-halogen lamp that
illuminates a white-spot on the AAT dome.  The fibre intensity is traced
in a two stage process. First the fibre peaks are
identified and fitted approximately using a quadratic fit to the 3
pixels around each peak (this gives positions accurate at the
$\sim0.1$ pixel level).  Then as a second stage (newly implemented for
the SAMI pipeline) we implement an algorithm that assumes a Gaussian
fibre profile (a good approximation to SAMI fibres in AAOmega) and
fits five Gaussians (the central one and two either side) to precisely
determine both the centre and width (to be used later for optimal
extraction) of the fibre profile.  When fitting we integrate the
Gaussian model across each pixel.  A robust 4th order polynominal is
then fit to both the tramline and width of each fibre as a function of spectral pixel.  The multiple
Gaussians are required as the close packing of fibres causes the wings of their light profiles to
overlap at the $\sim5-10$ percent level so that adjacent fibres
influence the measured position and centre of the fibre in question \citep{sharp10a}.

\begin{figure}
\includegraphics[width=80mm]{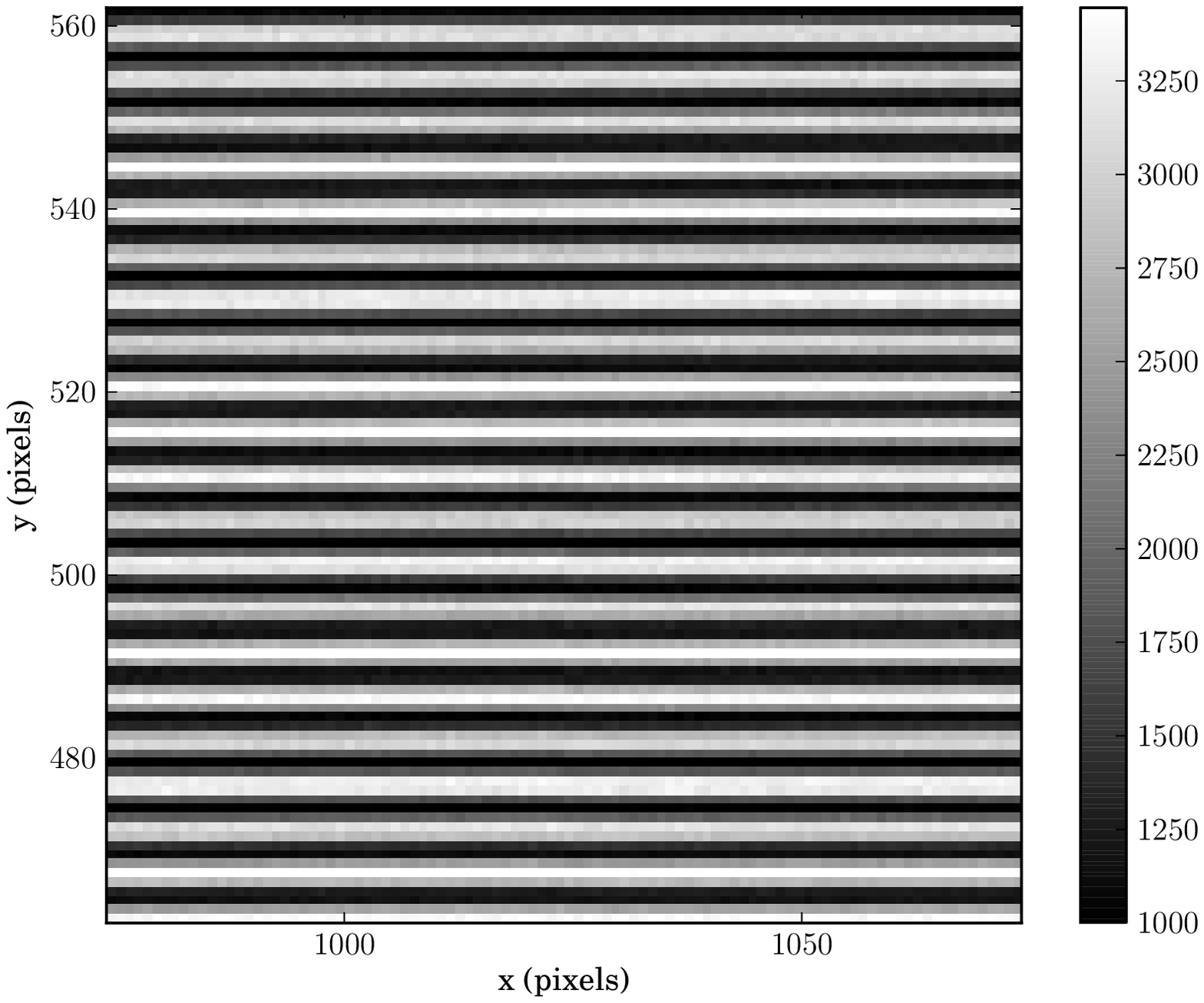}\\
\includegraphics[width=80mm]{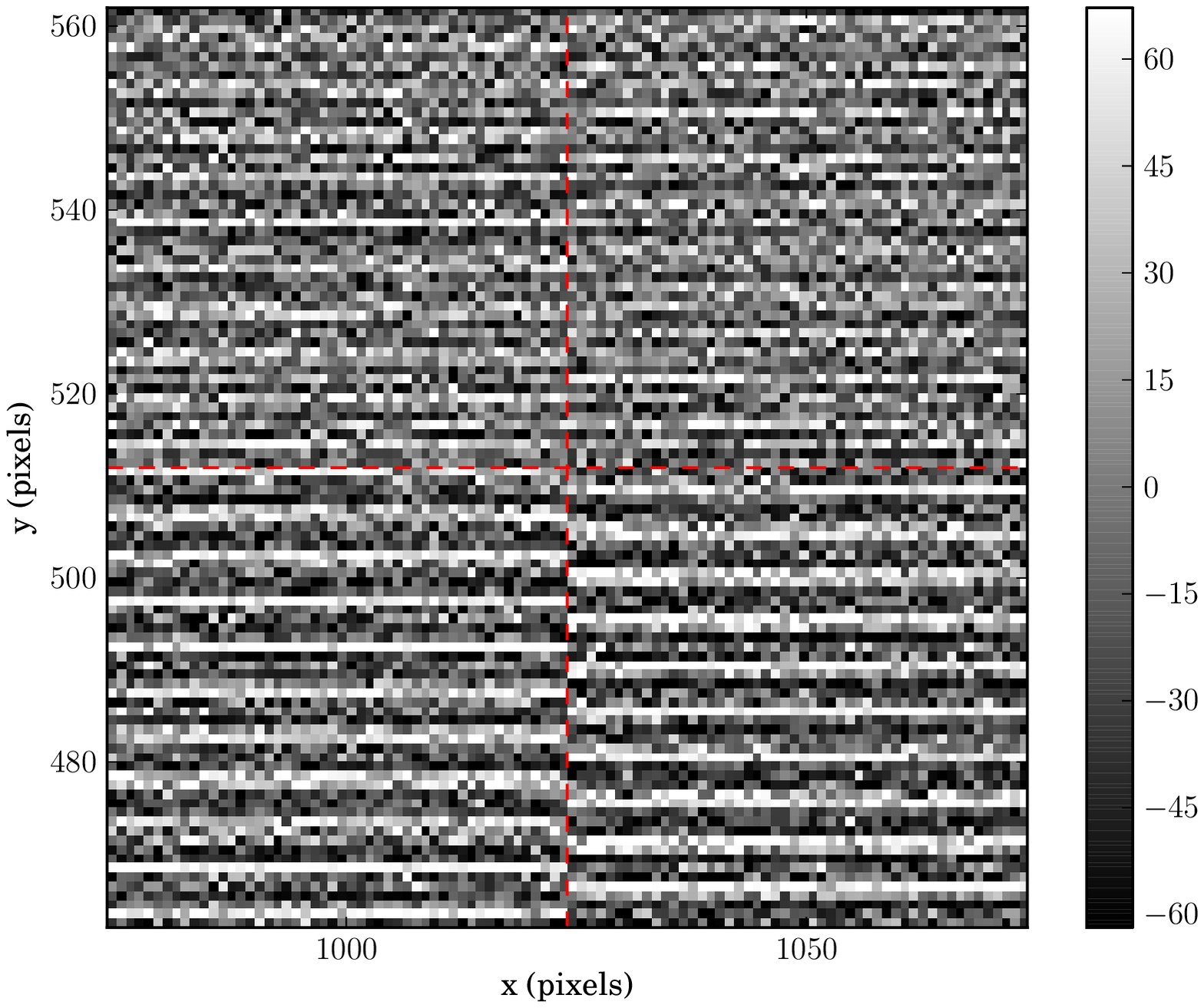}\\
\includegraphics[width=80mm]{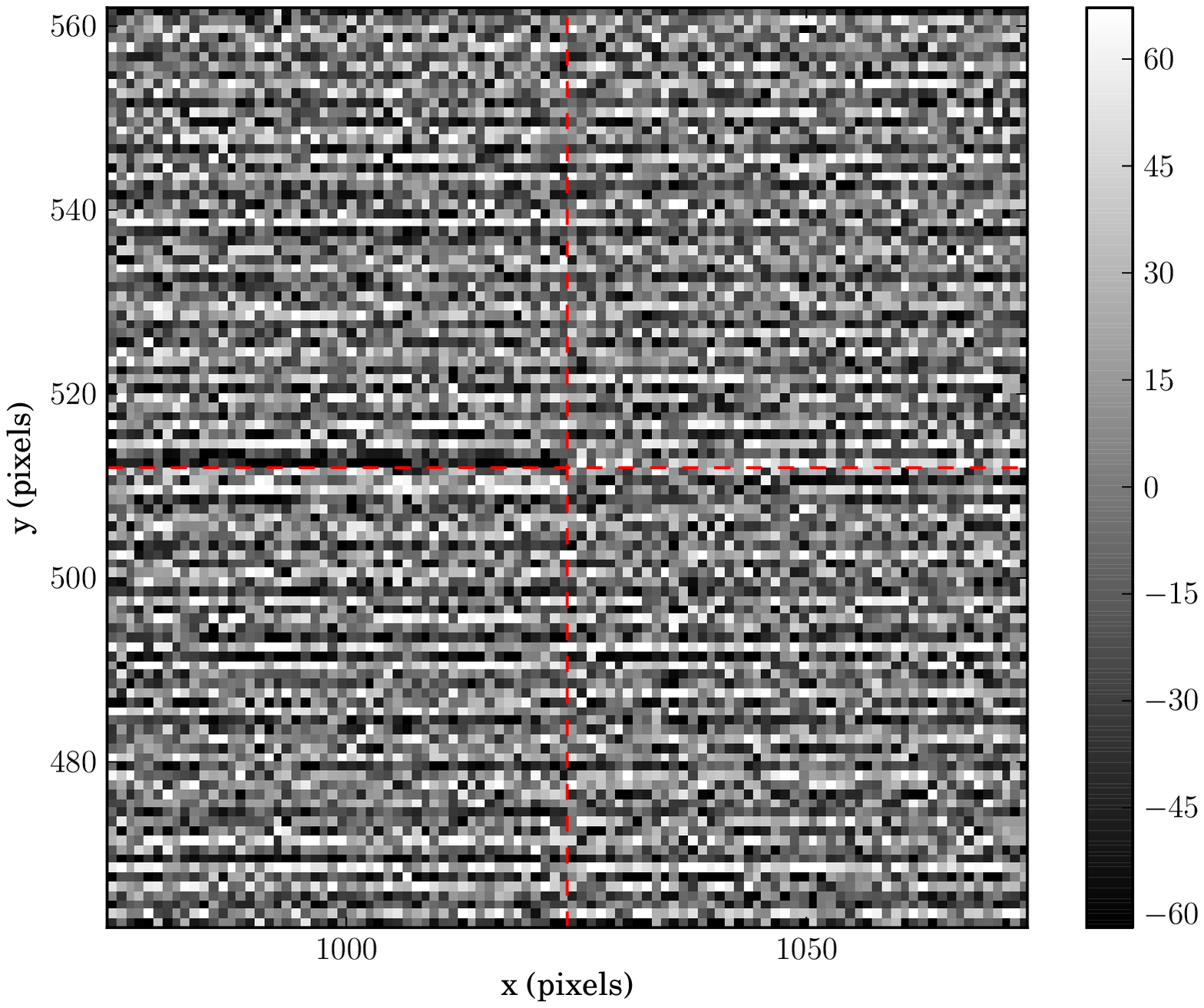}
\caption{A small region of a SAMI fibre flat field image (top), along
  with residuals after extraction (middle and bottom).  The residuals are shown
  without (middle) and with (bottom) correcting for lithography errors.
  The red dashed lines indicate the CCD lithography boundaries where
  we would expect discontinuities in the fibre trace.}
\label{fig:litho_im}
\end{figure}

\begin{figure}
\includegraphics[width=95mm]{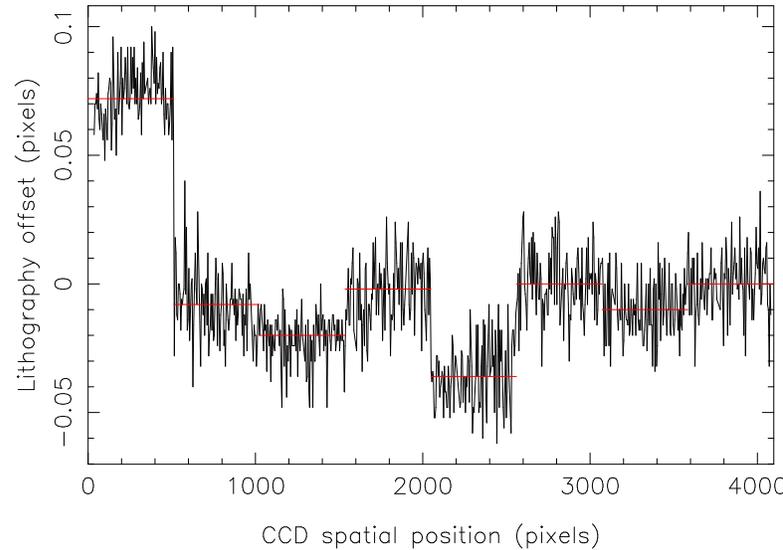}
\caption{CCD lithography errors measured during fibre tracing.  The
  black line shows the offsets measured for each fibre and the red
  lines show the median measured offset in each 512 pixel block.}
\label{fig:litho_errors}
\end{figure}

Our approach allows two systematic effects present in the fibre traces to be addressed.
The first is that the AAOmega CCDs, 2$\times$4k E2V detectors, are manufactured using a lithography
mask of 1024$\times$512 pixels. There are errors in the relative
positioning of the mask on different parts of the detector, causing
discontinuties in the fibre traces between pixel 1024 and 1025 (in the
spectral direction).
Adding a step when fitting the fibre trace allows
us to accurately measure these lithography alignment errors, we find the
typical uncertainty on an individual fibre is $\simeq0.01$ pixels.
Because there are $\sim100$ fibres per lithography block, the mean error
per block can be derived at $\sim0.001$ pixel precision.  The worst
lithography errors were found to be $\simeq0.07$ pixels, but most are
substantially less than that $\simeq0.01-0.02$ pixels.  A small region
of a fibre flat field is shown in Figure~\ref{fig:litho_im}.
The residual image (the difference between the input and model flat field data)
without correction for lithography errors (middle panel)
shows clear discontinuities that are removed after the fibre traces
are corrected for this error (lower panel).  In
Figure~\ref{fig:litho_errors} we plot the lithography errors measured
for each fibre and the median values in each block for a typical fibre
flat field from the AAOmega red arm. 

A second systematic discovered was a slow shift in position of the
fibre traces of approximately 0.02-0.03 pixels per hour.  Investigation
showed this to be due to a time-varying gravitational torque on the
AAOmega cameras caused by the slow boiling off of the liquid nitrogen in
the dewers attached to the cameras.  Comparison of tram-line maps to
the residuals from extracted data frames suggests that our typical
total tramline map error is $\simeq0.02$ pixels, and never worse than
$0.05$ pixels.

\begin{figure}
\includegraphics[width=90mm]{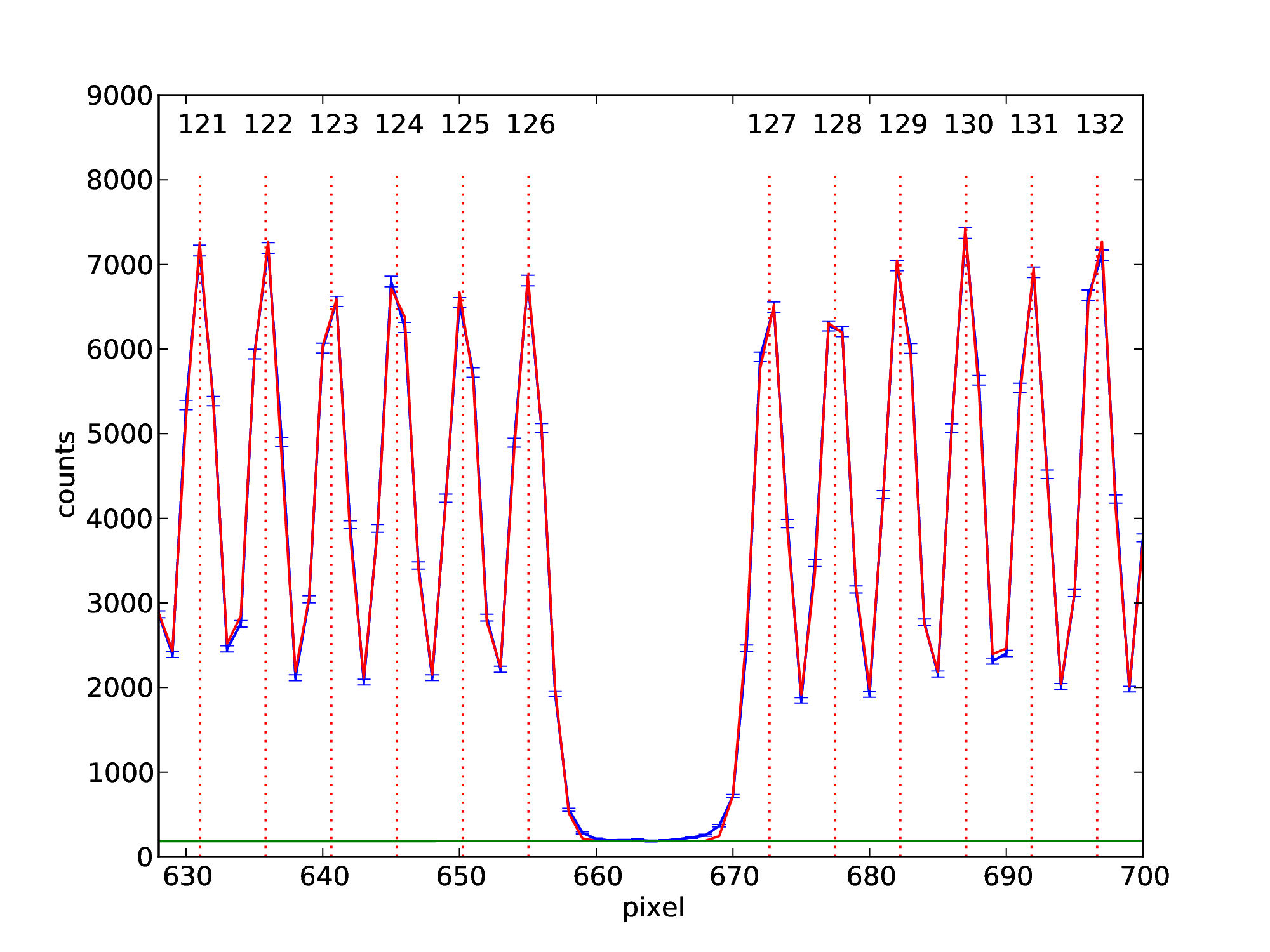}
\caption{Example fibre profile data (blue line with errors) from a
  fibre flat field frame.  Over-plotted (red) is the best fit multi-Gaussian
  model from our optimal extraction.  The smooth scattered
  light model is shown in green.  The vertical dashed lines show the
  expected centres of each fibre from the tramline mapping process,
  with fibre number along the slit labelled above.  The gap at pixel
  $\simeq665$ between fibres 126 and 127 is due to the separation
  between individual slitlets each containing 63 fibres in the SAMI slit.}
\label{fig:optex}
\end{figure}

After measuring the tramline map and fibre profile widths, the next
stage is extraction of the flux from the 2D image to generate a 1D
spectrum for each fibre. An optimal extraction \citep{sharp10a} is
performed to fit the flux amplitudes perpendicular to the dispersion
axis. Gaussian profiles are fit, holding the centre and width
constant (based on the tramline and fibre width maps measured
above) and fitting all 819 fibres simultaneously.  At the same time a
B-spline is fit to model the smooth scattered light.  This is typically
8th order, so that a total of 827 parameters are fit at
once for each CCD column (4096 pixels).  To enforce smoothness on the scattered
light model, the fit is done in two passes.  On the first pass all 819
fibres plus 8 scattered light parameters are fit \citep[as outlined
in][]{sharp10a}.  Then the scattered light model is
smoothed across columns and the resulting 2D image is subtracted from
the data frame.  The second pass is then done on the subtracted frame,
fitting only the 819 fibre amplitudes, without any scattered light
model.  An example extraction fit is shown in Figure~\ref{fig:optex}.

Following extraction, the 1D spectra are divided by an extracted and
normalised 1D fibre flat field spectrum, which removes residual
fibre-to-fibre variations in spectral response.  These do not correct
for total throughput as the  illumination is not sufficiently uniform
across the flat field and so no transmission calibration between the
fibres is possible at this stage \citep{sharp13}.  Wavelength calibration is performed using
standard arc-lamps (CuAr). Emission lines are identified in extracted 1D
spectra and matched to line-lists with a 3$^\mathrm{rd}$ order
polynomial for each fibre solution.
A secondary wavelength calibration is performed in the red arm by measuring the positions of several sky emission lines and fitting a quadratic to the residuals relative to their known wavelengths. This modification corrects for small shifts introduces due to due to difference in the feed angle of sky and calibration illumination. It is found to improve sky subtraction accuracy while not significantly modifying the wavelength solution.
This correction cannot be applied in the blue arm, where only the 5577-\AA sky line is in the spectral range observed.

The relative throughput between all fibres is established via
measurement of several isolated night sky emission
lines. The requirements for relative transmission
calibration are outlined by \citet{sharp13}.
In occasional cases the sky lines are affected by cosmic rays and other artefacts.
To minimise the effect of this issue, we use the median throughput value for each fibre across all
observations of a field on a single night

The 26 individual sky
fibres within the SAMI spectrograph slit are configured to blank sky
positions in each observation field.  Once throughput calibrated, a
master sky spectrum is generated by stacking the individual sky fibres
with each frame before subtracting the sky. 
The Principal Components Analysis technique \citep{sharp10b} cannot be
used due to the common spectral features in most SAMI Galaxy Survey targets,
a consequence of the limited redshift range for the survey and the
coherence of spectral structure within each galaxy. 

The individual fibre spectrum processing with
\texttt{2dfdr} is now complete and the Row-Stacked-Spectra (RSS) frames
are passed to an external processing suite to flux calibrate and
correct for differential atmospheric refraction (DAR) and dispersion before aligning
and mosaicking individual objects to produce datacubes. 

\section{Flux calibration and telluric absorption correction}
\label{fluxcal}
The flux calibration for each set of galaxy observations has two parts. First, a spectrophotometric standard star -- typically observed on the same night as the galaxy observations -- is used to correct for the large-scale wavelength dependence of the instrumental transmission. A secondary standard star -- observed simultaneously with the galaxies -- is then used to correct the telluric absorption bands. A full analysis of the flux calibration accuracy achieved for the SAMI Galaxy Survey Early Data Release is given by \citet{allen14b}. Below we briefly describe the principles adopted.

\subsection{Extracting total stellar spectra}
\label{starspec}
In principle, each stage in the flux calibration can be performed by multiplying the observed galaxy spectra by the ratio of the true flux to the observed flux of a standard star. For IFU observations with a filling factor less than 100 per cent, we must also account for the light that falls between the fibres. Crucially, the fraction of light lost in this manner is a function of wavelength, as a result of  atmospheric dispersion and the dependence of seeing on wavelength. We make the correction by fitting the observed stellar flux across the hexabundle with a model PSF, including the atmospheric effects. The model PSF as a function of $x$, $y$ position (in arcseconds) and wavelength, $\lambda$, takes the form of a \citet{moffat69} profile:
\begin{align}
	p(x, y, &\lambda | x_{\rm ref}, y_{\rm ref}, \lambda_{\rm ref},\alpha_{\rm ref}, \beta) =  \nonumber \\
	&\frac{\beta - 1}{\pi \alpha(\lambda)}
	\left(1 + \left( \frac{(x - x_0(\lambda))^2 + (y - y_0(\lambda))^2}{(\alpha(\lambda))^2} \right) 
	\right)^{-\beta}, 
\end{align}
where $x_{\rm ref}$, $y_{\rm ref}$ and $\alpha_{\rm ref}$ give the position and size at
an arbitrary reference wavelength $\lambda_{\rm ref}$. The dependence on wavelength is given by
\begin{align}
	x_0(\lambda) &= x_{\rm ref} + 206250 \left( n(\lambda) - n(\lambda_{\rm ref}) \right) \tan(ZD) \sin(\phi), \nonumber \\
	y_0(\lambda) &= y_{\rm ref} + 206250 \left( n(\lambda) - n(\lambda_{\rm ref}) \right) \tan(ZD) \cos(\phi), \nonumber \\
	\alpha(\lambda) &= \left( \frac{\lambda}{\lambda_{ref}} \right)^{-0.2} \alpha_{\rm ref},
\end{align}
where $ZD$ is the zenith distance and $\phi$ is the parallactic angle. The refractive index of air as a function of wavelength, $n(\lambda)$, is given by equations 1--3 of \citet{filippenko82}, which are in turn functions of temperature, pressure and water vapour pressure.

To constrain the free parameters, the observed wavelength range in each CCD is divided into 20 chunks of 100 pixels each, having discarded the 24 pixels at the beginning and end of each CCD. The observed counts in each fibre are summed within each of these wavelength ranges, and the model PSF is fitted to this summed data. The data from both CCDs are fitted simultaneously, in order to have a wide wavelength range to constrain the atmospheric dispersion effects. During this fit, the parameters $x_{\rm ref}$, $y_{\rm ref}$, $\alpha_{\rm ref}$, $\beta$ and $ZD$ are allowed to vary along with the total flux in each wavelength range, while $\phi$ and the atmospheric parameters are fixed to their measured values\footnote{The zenith distance dependance is largely degenerate with atmospheric pressure. Fitting accuracy was found to improve marginally if minor variations in $ZD$ were allowed as a free parameter of the model.}.
For SAMI observations under typical observing conditions the uncorrected atmospheric dispersion between 3900\AA\ and 7270\AA\ is of the order 1\arcsec, $\sim$60\% of a fibre core diameter. After correction the mean offset between these two wavelengths, averaged over a series of observations, is found to be 0\farcs14. At less than 10\% of the fibre core diameter, this is deemed to be at the limit of measurement.

With the PSF parameters set, the final step in extracting the spectrum is to fit for the overall flux, i.e.\ the scaling of the PSF, in each wavelength pixel. A uniform background level is also fit in each wavelength pixel, to allow for residual errors in the sky subtraction. Each wavelength pixel is fit independently. The result is a spectrum recording the total number of CCD counts that would have been observed, if the filling factor of the IFU was 100 per cent.

\subsection{Primary flux calibration}
\label{sec:prifluxcal}
The spectrum for the primary standard (typically multiple standard stars are observed, at a range of airmasses)
is then corrected for atmospheric extinction using the default Siding Spring Observatory (SSO) extinction curve (scaled for airmass) and rebinned to match the wavelength sampling of the
reference spectrum, given in units of erg\,s$^{-1}$\,cm$^{-2}$\,\AA$^{-1}$. The
ratio of the observed and reference spectra is then used to infer the
wavelength-dependent scaling needed to calibrate the data in an absolute
sense, or, equivalently, the instrumental response function, $R(\lambda)$.

To minimise the effect of noise in the standard star observations, and of small-scale mismatches between the observed and template spectra, the measured ratio is smoothed by fitting a spline function. Approximately eight spline knots are used within each arm, although extra knots are inserted around 5500\,\AA\ where $R(\lambda)$ shows a sharp turnover due to the dichroic cut off, and knots that lie within telluric bands are removed.

After smoothing, multiple observations of each standard star from a given night are combined to produce the final calibration. The agreement between different observations is very good in terms of the shape of $R(\lambda)$, but there is some variation in the overall scaling with a standard deviation of 11.1\% found in the available year-1 data set. This variation is driven by a combination of changes in atmospheric transmission and a weak degeneracy between the scaling and the FWHM in fitting the PSF. To remove the effect of any outliers with discrepant normalisations, the individual measurements of $R(\lambda)$ in each arm are re-scaled such that their value in the centre of the wavelength range is equal to the median of these values across the set of observations. The re-scaled measurements of $R(\lambda)$ are then combined using a simple mean. The response function is then applied to each science frame after correction with the airmass-scaled SSO extinction curve. Details of the accuracy of the calibration process, with SAMI datacubes referenced to ancillary data (e.g., SDSS photometry and spectroscopy) are given for the SAMI Galaxy Survey Early Data Release by \citet{allen14b}.

\subsection{Telluric absorption correction}
\label{sec:secfluxcal}
The Fraunhofer B-band, a telluric absorption feature due to atmospheric O$_{2}$ at 6867\,\AA, falls within the red portion of the SAMI Galaxy Survey spectra and requires correction. Because the telluric band strength is variable and strongly correlated with airmass, the correction is best derived from data contemporary with the science observations. For every science observation, one of the thirteen SAMI bundles is allocated to
a secondary standard star. These secondary standards are colour
selected to be F sub dwarfs, which have relatively flat and featureless
spectral shapes, allowing simple modelling with either synthetic or empirical
template spectra. The total spectrum for each secondary standard is extracted using
the process described in \S\ref{starspec}, after the RSS data have been flux calibrated as per \S\ref{sec:prifluxcal}.
The secondary
standards are used to correct for telluric absorption. As a byproduct of
this process, the secondaries also provide per-dataframe information about the
seeing, including an empirical measurement of DAR. This information is used in the later alignment and drizzling stages.

The spectrum of a F sub dwarf is sufficiently smooth around the regions of telluric absorption (6850--6960\,\AA\ and 7130--7360\,\AA) that it can be modelled with a  simple straight line fit. The fit is performed using all data outside of the telluric regions and redwards of H$\alpha$. The telluric absorption is then given by the ratio of the extracted spectrum to the straight line fit; all galaxy spectra in the frame are divided by this ratio, with the correction applied only in the vicinity of the telluric feature.

\subsection{Absolute flux calibration}
The flux calibration procedure assumes no change in atmospheric conditions (other than for the telluric absorption features) between the observations of the spectrophotometric standard star and the galaxy field. In practice, some variation can occur, which to first order produces a wavelength-independent scaling of the observed flux. To correct for this scaling, the spectrum of the secondary standard star is extracted from the combined datacubes using the same procedure as described in
Section~\ref{sec:prifluxcal}, and converted to a $g$-band magnitude by integrating across the SDSS filter curve \citep{Hogg02}. This measurement is compared to the available photometry to find the appropriate scaling, which was then applied to all objects in the field.

\section{Resampling to a Cartesian grid}
\label{drizzle}
The SAMI hexabundle format imposes two restrictions on image reconstruction which are shown graphically in Figure~\ref{layout}. While the positions of the individual fibre cores that make up each IFU are well defined ($\pm1\,\mu$m relative errors, $\sim$1\% of a fibre diameter), they are on an irregular grid. Secondly, each hexabundle IFU has a 73\% fill-factor for a single observation. The full image profile is recovered through a conventional series of dithered observation with the telescope offset to fill in the gaps in target coverage. Dithering generates a series of misaligned data frames which possess a well registered but non-common geometry. In principle, one can retain the data in this native format, but the exigencies of scientific analysis typically dictate that data should be rebinned, often to a common Cartesian output grid. This facilitates straightforward stacking of dithered frames to allow outlier rejection and accrual of S/N as well as easy visualisation of resulting data products, and significantly simplifies many subsequent analysis steps.

A simple option for resampling is to perform a nearest-neighbour interpolation of the hexabundle input data onto a regular Cartesian grid. An immediate effect of such an operation is to effectively convolve the intrinsic resolution of the spatial data with a kernel corresponding to the chosen output spaxel size, blurring the image. Secondly, such a resampling introduces a complex covariance between output spaxels which overlap more than one input fibre core. Both issues can be minimised by selecting a fine pitch for the output spaxel grid, but this produces a bloated data format, containing significant redundant information, and ultimately reduces S/N due to the excessive oversampling. The problem is well documented, and a solution clearly defined by \citet{fruchter02} through their introduction of the {\it Drizzle} algorithm, initially conceived to resample high-resolution imagining from $HST$/WPFC2, which sub-critically samples the PSF from $HST$.

\begin{figure*}
\begin{center}
\includegraphics[width=150mm]{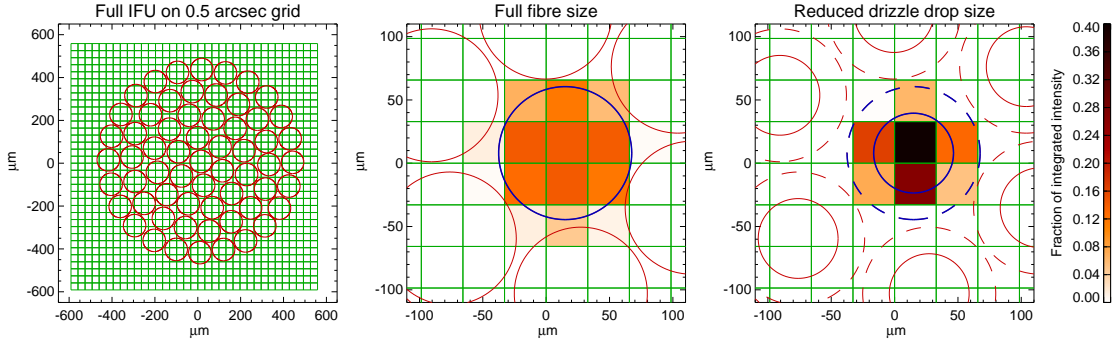}
\caption{\label{drizzleexample} An example of the SAMI implementation of the drizzle algorithm \citep{fruchter02} shows the redistribution of flux for a SAMI IFU bundle onto a regular Cartesian output grid. The 0\farcs5 output spaxel grid is chosen to encompass the entire IFU dither pattern for a series of observations (left), the regular output grid is shown below the fibre bundle footprint for a single dither position. For each input fibre core, each output spaxel receives a portion of the input flux (centre). As we show in \S\ref{dropsize}, loss of spatial resolution can be minimised by distributing the flux from each input fibre core as if it had a reduced the diameter (right) with the flux distributed over a smaller number of output spaxels and at higher intensity in each.}
\end{center}
\end{figure*}

For a detailed description of the drizzle algorithm, the reader is directed to \citet{fruchter02}. Briefly, considering each input fibre core in turn, and for any given input geometry, one can calculate the overlap area of the input fibre core with each element of a predefined regular grid of output spaxels. The fractional area of an input fibre core covering each output spaxel dictates how the flux should be redistributed to each output spaxel (Figure~\ref{drizzleexample}). This fractional area provides a {\it weight} for each output spaxel which represents the relative exposure of each output spaxel.

\begin{figure*}
\begin{center}
\includegraphics[width=55mm]{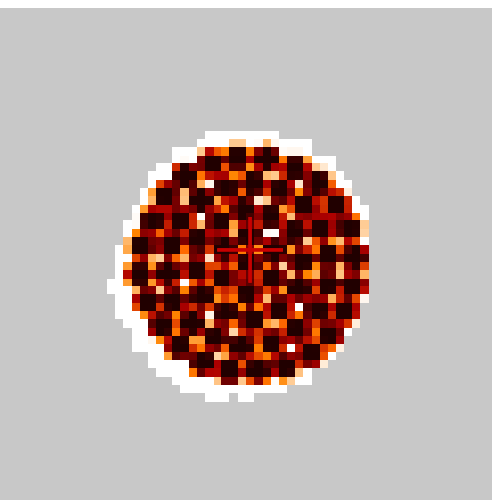}
\includegraphics[width=55mm]{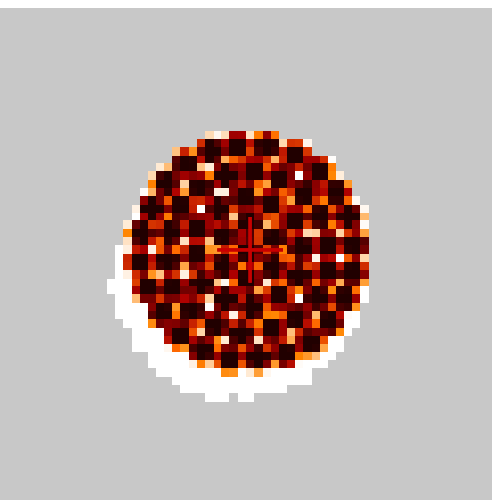}
\includegraphics[width=55mm]{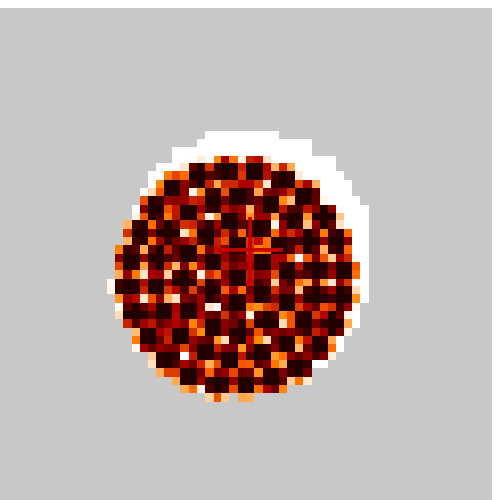}
\caption{\label{weight} Weight maps are shown for three datacubes from an aligned dithered data set. Gray spaxels have no coverage in the current output geometry, while the white-to-black scaling indicates weights between 0 and 1. The fibre core centres are visible as spaxels with saturated (black) weights. Note, the large white borders to each cube show zones of zero overlap between an individual image and the overall image stack. Weight maps can be interpreted as the relative exposure-time of each spaxel of the image. The output spaxel scale is set to 0\farcs5 which means that while most output spaxels have some fractional contribution from one or more SAMI fibres, some spaxels inside individual cubes have no contributing data.}
\end{center}
\end{figure*}

Output spaxels that do not sit within any input fibre cores will be assigned a weight of zero. Output spaxels that fall on the border between multiple input cores will have a weighted contribution from each core, but the total weight for such spaxels will not exceed unity (since by definition, an output element cannot be completely inside one input element if it is also part of other input elements). To place dithered data onto the regularised grid, one merely recalculates the overlaps after perturbing the baseline position (by the known telescope offset) of the input fibre cores relative to the initial reference position. Figure~\ref{weight} presents an example of the relative weight mappings for three observation that make up part of a dither set.

For isolated output spaxels, this process conserves total flux but does introduce covariance, an issue we return to in \S\ref{neglectingcovar}. In this regime, the associated error information for each spaxel can be redistributed such that the global signal-to-noise is preserved, i.e., the input variance is simply weighted by the square of the weight map.

\subsection{Weight cubes}
\label{weightcubes}
The weight maps derived from the drizzle resampling trace the effective exposure time for each output spaxel. By design, not all output spaxels have the same effective exposure time. Those that are not completely covered by one or more IFU fibre cores will have a reduced intensity proportional to the fractional area of the output spaxel covered by input fibre cores. Depending on the output spaxel size chosen, some spaxels may have an effective exposure time of near zero even though they reside within the physical boundary of the IFU bundle. For this reason, a reconstructed image of a source within the drizzle resampled datacube will exhibit an intensity structure dominated by the relative weights of its component spaxels, largely obscuring the underlying source structure.

\begin{figure}
\begin{center}
\includegraphics[width=85mm]{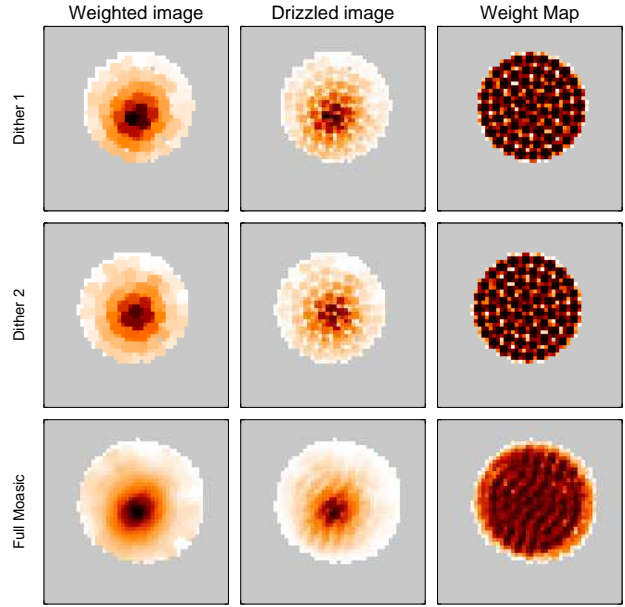}
\caption{\label{weightedimages}
The effect of the weight map is illustrated with three mosaics of a SAMI galaxy. The two upper rows each show a single dither position observation for the galaxy.
The bottom row shows the reconstructed image after seven dither positions are combined.
The first column shows the reconstructed image after division by the weight map, showing a constant effective exposure time for all spaxels. The second column shows the raw output from the drizzle process where the intensity distribution is heavily modulated by the different effective exposure times of each output spaxel due to the partial coverage by input fibre core. The third column shows the weight map, the effective exposure time for each spaxel.}
\end{center}
\end{figure}

To overcome this effect, the resampled datacubes (and their associated variance arrays) are stored with the weight map normalisation pre-applied (i.e., mosaic cubes are divided by their weight map in the default data product). This normalises the effective exposure time across the datacube and recovers the underlying source structure.  This data format means most users may safely ignore the weight maps for most applications. Provided the weight map is propagated alongside each data frame, the process of conversion between the two formats is reversible and deterministic. Examples of three weight maps for a three-point dithered observation are shown in Figure~\ref{weight} and the imprint of variations in relative exposure across a resampled observation is shown in Figure~\ref{weightedimages}.

As dithered observations are combined, the relative weighting of all spaxels will approach uniformity across the mosaic and the relative exposure structure will be removed from the datacube. The practical implication of this is :\\
{\it Local properties }of a source, such as surface brightness of emission features or light profile fitting, should use the default data format with the weighting already applied to the datacubes.\\
{\it Global properties }of a source, such as integrated H$\alpha$ flux or continuum intensity, should pay strict attention to the weight map values to avoid erroneously including additional signal, particularly from the edges of the mosaic, which are highly scaled up from low relative exposure times with respect to the central regions of the cube.

When combining data, cubes are first multiplied by their weight maps, the data and individual weight maps summed, and the summed data finally divided by the new weight map to preserve the correct flux calibration.

\subsection{Dither frame alignment}
The relative offsets between each dithered dataset must be known before they can be drizzled 
to a common output spaxel grid and combined. In principle, these offsets can be obtained from 
telescope pointing and offset information available in each RSS frame header. 
In practise this information does not provide the required level of accuracy, 
particularly when data are taken with multiple source acquisitions (for example over multiple nights), a process which can affect 
the base pointing position. 

A number of methods were explored for alignment, largely treating each IFU individually. 
These included a simple centroid fitting, cross correlation of reconstructed IFS images and alignment using SDSS 
imaging data as a reference frame. It was found that all work remarkably well for most 
source types, but fail significantly for some classes of objects, such as galaxies 
with disturbed morphologies, interacting pairs or low surface brightness systems. 

In order to overcome this limitation, and to recover all the observed targets in a uniform manner, an alignment 
technique has been developed to simultaneously estimate the dither patterns for all IFUs in a given observation. The principle treats each exposure as an image of the sky, and estimates the best-fitting coordinate transformation to 
align each RSS frame to a reference RSS frame (typically the first observation in a sequence). In this way, even if one (or more) IFUs cannot be used 
for the estimate of the coordinate transformation (for example due to a disturbed morphology system which provides unstable alignment results), the best-fitting solution allows us to recover the dither pattern for the entire frame. This goal is achieved in three steps. 

First, a 2D Gaussian is fitted to an intensity map for each IFU (obtained by collapsing the cube along the wavelength axis) 
in order to recover the centroid (i.e., peak signal) positions. Assuming that the absolute position of the peak emission on the sky 
does not vary between exposures, the centroid coordinates can be used to estimate the best-fitting coordinate transformation 
necessary to align all IFUs on a reference frame. 

Second, the best-fitting coordinate transformation is computed using a Python implementation
of the IRAF\footnote{IRAF is distributed by the National Optical Astronomy Observatory, which 
is operated by the Association of Universities for Research in Astronomy (AURA) under cooperative agreement with the National Science Foundation.}\textsc{geomap} task.
We allow for a combination of shifts in the $x$ and $y$ directions, a rotation and a common plate-scale change in the $x$ and $y$ directions. 
Specifically, the coordinate transformation has the following functional form:
\begin{align}
x_{ref} &= x_{shift} + \left(\Delta \times x_{in} \cos{\theta} \right) + \left(\Delta \times y_{in} \sin{\theta} \right), \nonumber\\
y_{ref} &= y_{shift} - \left(\Delta \times x_{in} \sin{\theta} \right) + \left(\Delta \times y_{in} \cos{\theta} \right)
\end{align}
where $x_{ref}$ and $y_{ref}$ are the centroid positions in the reference frame, $x_{in}$ and $y_{in}$ are the centroid positions of the plate to be aligned, 
$x_{shift}$ and $y_{shift}$ are the rigid shifts in the x and y directions, $\Delta$ is the magnification factor and $\theta$ is the rotation angle.
A 2$\sigma$ clipping technique is applied to remove those IFUs (3 on average) for which the 2D Gaussian fit is unstable. 
The mean values of shift, magnification and rotation angle found for our data are $\sim$35\,$\mu$m, $\sim$10$^{-4}$
and $\sim$0.014\,degrees, respectively.

Third, we use an implementation of the IRAF \textsc{geoxytran} task to apply the coordinate transformation to the central fiber of each IFU and determine its position on the reference plate. For each IFU the relative offset between the two exposures 
is then given by the difference in the positions of the central fibers.  While the rotation term introduces a significant translation in base [$x$,$y$] position between observations, the magnitude is sufficiently small that no accounting is made for rotation of fibre positions within a bundle (the correction for internal rotation would be typically of the order 0.2\,$\mu$m, less than 1\% of a fibre core diameter and smaller than the relative positional uncertainty of fibre cores within each bundle).

In data analysed to date, the typical rms for the final dither solution is $\sim$12$\,\mu$m $\pm$8$\,\mu$m (i.e., 1/9$^\mathrm{th}$ of the fibre size).
These values are found to be consistent with those obtained with the single IFU methods for stable cases
indicating the technique is providing a high accuracy and stable solution for all IFUs across a given SAMI observation.

\subsection{Simple summation combination}
Once relative alignment has been determined, dithered observation data sets can be combined. We start by assuming $n$ datacubes are available from a dithered set. Each datacube $C_n$ is an array of [$x$,$y$] spaxels (which for compactness we shall denote by $r$, with the assumption that $r$ is defined in a reference frame with all cubes aligned). We also assume, as is the default for SAMI data, that the datacubes are stored with division by the weight map pre-applied (\S\ref{weightcubes}). Each cube also has $\lambda$ wavelength elements that, due to atmospheric dispersion, are offset with respect one and other. This misalignment is corrected by recomputing the drizzle mapping solution as a function of wavelength at regular intervals. The interval is chosen such that the accumulated dispersion misalignment is never more than 10$^{\rm th}$ of a spaxel.
datacubes need not be a common size in [$x$,$y$], although the spaxel scale of the cubes to be combined must be constant\footnote{In principle the inputs to the drizzle resampling step need not all be at a common scale, provided the correct geometry is supplied to the drizzle code.} and the region of union between cubes well defined. Each cube has an associated variance map $V_n$ and map of the relative spaxel weights $W_n$.

We define an output datacube, $C_{out}$, and its associated variance array, $V_{out}$ and weight map, $W_{out}$, such that

\begin{align}
\label{sum}
W_{out}(r,\lambda) = &\sum_n W_n(r,\lambda),\\
C_{out}(r,\lambda) = \frac{1}{W_{out}(r,\lambda)} &\sum_n C_n(r,\lambda) W_n(r,\lambda),\\
V_{out}(r,\lambda) = \frac{1}{W_{out}^2(r,\lambda)} &\sum_n V_n(r,\lambda) W_n^2(r,\lambda).
\end{align}

The default values of $C_n$ must be multiplied by their weight maps in order to restore the true relative exposure time structure to the input data before summation. After summation, the output cube is divided by the new weight map to remove the relative exposure time structure from the final data product.
Not applying the weight map would leave the cube scaled for differences in the effective exposure times for each spaxel and would not correctly propagate observed flux.
For many spaxels the weight will eventually exceed unity as multiple exposures are summed together. This merely indicates that the spaxel is fully sampled at a higher signal-to-noise than would be achieved for unit exposure time, due to multiple overlapping observations. While these relative exposure values must be carefully propagated in the image header, they present few problems. Arbitrary renormalization of the weight map is permissible, provided the effective exposure time of the associated datacube is also renormalised by the same factor. This preserves the correct surface brightness for the datacube.

\subsection{Outlier rejection}
Cosmic rays and other defects
will remain in the reduced datacubes at some level. Additionally, bad pixels from the CCD will contain no data and hence will be missing/flagged in the datacubes. Clipping flagged values is as simple as removing them from summation and reducing the corresponding value of $N$ for each $(r,\lambda)$ in Eq.~\ref{sum}. The key is accurate flagging.

To remove outliers from the summation of pixel values a clipped mean is preferable because it simplifies variance propagation.
To generate the clipping flags, we work on each spaxel and spectral pixel of the output cube $(r,\lambda)$ independently, noting that there will be $N$ intensity values at each output pixel, one from each input datacube.  Note also that some input values will have weight $W_n(r,\lambda)$=0 due to the nature of the dithered observation. We generate the working vector, $m(n)$, and its associated variance array such that 

\begin{eqnarray}
m(n)=[C_0(r,\lambda | W_0\ne0),...,C_n(r,\lambda| W_n\ne0)]
\end{eqnarray}

At this point a simple sigma clipping rejection, with a modest threshold, will fail due to the finite fibre footprints which sample different parts of each source due to the dithered observation stratergy.
Put simply, in comparing the dithered input spectra at each spaxel, one is not directly comparing like-with-like.

For all real sources, which lack spatial discontinuities on the scale of SAMI data due to the seeing profile, the first order difference will simply be intensity. There will be no appreciable change in spectral shape across the output spaxels. Therefore, if each input spectrum is first normalised to unity prior to construction of the vector $m(n)$ the sigma-clipping outlier rejection flags only significantly deviant pixels (a 5$\sigma$-clipping threshold is used). With the errant pixels now cleanly identified, the remaining good pixels are used, without normalisation, in the summation equations of Eq.~\ref{sum}. In this manner, highly discrepant data points are removed from the summation, while still retaining the correct intensity information from the dithered data set.

\subsection{Confirming variance propagation accuracy}
The default data product presented above generates a variance array to track signal-to-noise for each input fibre spectrum as it is added to the dithered output mosaic. This simple data product fails to account for the significant correlation introduced between adjacent spaxels.
By definition the covariance matrix is given by
\begin{eqnarray} 
\Sigma(i,j)=\mathbb{E}\left[ \left( x_i - \mathbb{E}[x_i] \right) \cdot \left(x_j-\mathbb{E}[x_j]\right)\right],
\end{eqnarray}
where $\mathbb{E}(x_i)$ is the expectation value of the data in spaxel $i$.
%
%
Consider two adjacent output spaxels of the data cube, $i$ \& $j$ which sit beneath a common input fibre-core with $I_0\pm\sigma_0$ that contributes a fraction of flux to each output spaxel $\alpha_i$ \& $\alpha_j$. Then $\mathbb{E}\left(x_i-\mathbb{E}(x_i)\right)=\alpha_{i}\sigma_i$ and hence the variance term for spaxel $i$ is $\Sigma(i,i)=\sigma_i^2=\alpha^2_{i}\sigma_0^2$ and the covariance between output spaxels is $\Sigma(i,j)=\alpha_{i}\alpha_{j}\sigma_0^2$.

In order to test this error propagation model, a simulated dataset is generated. A dithered sequence of noise-free image frames are generated and drizzled onto the standard output grid. The input data is then duplicated and a Gaussian random noise field of known distribution is added. This second data set is also drizzled onto the standard output grid. If the associated noise properties of the image are correctly propagated by the error model, then a histogram of the difference between pixel intensity values for the noise-free and noisy data, scaled by the error array, will be a Gaussian distribution centred on zero with unit width. This is confirmed to be the case
, indicating accurate propagation of variance information from RSS frames to SAMI datacubes. This analysis confirms the correct propagation of the statistical errors recorded during the data reduction process. Underlying systematic variations in the data cubes, due principally to variations in observing conditions during survey operations, are assessed by \citet{allen14b}.

In essence, the drizzle process creates a number of identical copies of each input spectrum, each copy scaled by a value $\alpha_i$. Since duplication and scaling of a spectrum must retain the intrinsic noise properties, the associated error array is simply scaled by $\alpha_i$ to maintain the S/N in each copy of the spectrum. The noise properties of each output spaxel are correctly traced, but this simple model has made no accounting for the covariance between output spaxels, it merely correctly retains the noise properties of individual input spectra.


\subsection{Neglecting covariance}
\label{neglectingcovar}
Neglecting covariance between spaxels makes the assumption $\sigma_{ij}^2=0$ for $i\ne j$. With this assumption, on summation of the output error values for a dithered data set we no longer recover the input error, $\sigma_0$. As the $\alpha_i$ (with $0 \le \alpha_i \le 1$) sum to unity, $\sum(\alpha_i) = 1$, the quadrature sum is generally less than unity, $\sum(\alpha_i^2) \le 1$.
The S/N is correctly modelled within each individual output spectrum, but the noise level is underestimated across the full mosaic due to the lost covariance information.
While a crude fix for this approximation would be to rescale all the $\sigma_i^2$ by the factor $1/\sum(\alpha_i^2)$, a method for tracking the covariance is considered below.

\subsection{Tracking covariance}
The drizzle resampling methodology introduces unavoidable covariance between the output spaxels. For a detailed description of the problem, see \S7.1 of \citet{fruchter02}.
In principle the covariance information can be tracked for each output pixel. Indeed, the diagonal elements of the full covariance matrix (i.e. the variance) are already tracked in full. For an input fibre core diameter of 1\farcs6, and an output spaxel grid pitch of 0\farcs5, there are $\sim16$ output spaxels with non-zero covariance for each input fibre core (although only 4 -- 9 of these have significant covariance depending on the specific input/output geometry). On mosaicing dithered data of this format each output spaxel has a non-zero covariance only within a $5\times5$ spaxel grid of adjacent spaxels when adopting the default SAMI Galaxy Survey observing strategy (a seven point dithered mosaic with a dither pitch of 0\farcs72, to be considered in \S\ref{optimaldither}, onto 0\farcs5 spaxels with a 50\% drizzle drop size reduction to be presented in \S\ref{dropsize}). Explicit evaluation of the covariance between spaxels confirms there is no covariance outside of the $5\times5$ spaxel grid. However, even assuming these limited overlaps, providing the full covariance information in the output mosaic files would require a significant and largely unwarranted increase in the data volume and processing time for each observation.

When calculating and retaining the full covariance information, a datacube of  $\sim 1024$ Mb is generated, requiring a runtime of $\sim 100$ minutes on a standard dual core desktop machine --- a data volume and processing time that are prohibitively high for a large galaxy survey such as the SAMI Galaxy Survey. Both data volume and processing time can be significantly reduced by noting that the structure of the $5\times5$ covariance maps varies slowly with wavelength. By sampling at regular intervals in wavelength space and then interpolating between the measured values along the wavelength axis the full covariance matrix can be recovered with only minimal loss of information. The covariance cubes generated in this way are stored normalised to the variance (i.e., the central pixel of each 5$\times$5 covariance map has value 1.0) and require scaling by the variance to recover the correct magnitude. Further more, the covariance maps are generated from the overlap fractions of the input fibre footprints, and hence apply the flux and variance values before normalisation for relative exposure time with the weight maps.

As an example, consider the datacbe, $C[x,y,\lambda]$, and its associated variance and weight arrays, $V[x,y,\lambda]$ and $W[x,y,\lambda]$. Two adjacent spaxels, $A$ and $B$, will be covariant. The datacube contains the normalised spaxel values $C[x_A,y_A,\lambda]$ and $C[x_B,y_B,\lambda]$, each of which has an input flux and variance, prior to normalisation by the weight map, given by
\begin{eqnarray}
C'[x_n,y_n,\lambda] &= C[x_n,y_n,\lambda] \times W[x_n,y_n,\lambda]\\ 
V'[x_n,y_n,\lambda] &= V[x_n,y_n,\lambda] \times W[x_n,y_n,\lambda]^2
\end{eqnarray}
The covariance of spaxel $A$ with spaxel $B$ is contained within the covariance matrix, $Covar$, such that the covariance between $C'[x_A,y_A,\lambda]$ and $C'[x_B,y_B,\lambda]$ is given by
\begin{eqnarray}
Covar[x_A,y_A,x_B,y_B,\lambda] \times V[x_A,y_A,\lambda] \times W[x_A,y_A,\lambda]^2
\end{eqnarray}

\begin{figure}
\begin{center}
\includegraphics[width=70mm]{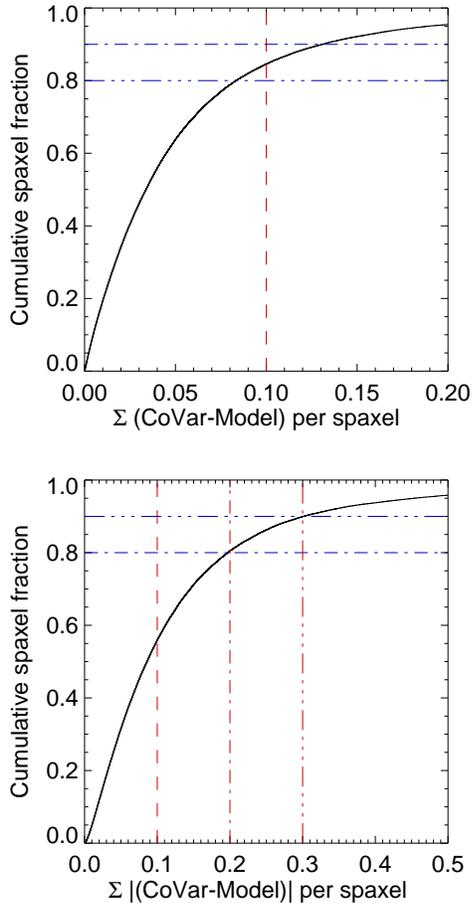}
\caption{\label{covarhist} The difference between the true covariance matrix and the modelled matrix is presented as the cumulative histogram of spaxels. The upper figure presents the cumulative histogram summing all under-estimates and over-estimates in the difference between the covariance model and the full matrix. The lower histogram considers the sum of absolute differences.}
\end{center}
\end{figure}

In addition to sampling at regular wavelength intervals, it is critical to finely sample the full covariance matrix (in the wavelength direction) either side of wavelength slices at which the applied atmospheric dispersion correction changes. The correction shifts the on-sky positions of the input fibre cores relative to their position on the CCD, and therefore alters the covariance between output spaxels. As a compromise between adequately sampling the full covariance matrix, and minimising the processing time required to produce a datacube and the volume required to store that cube we opted to sample the covariance matrix at 100 pixel intervals along the wavelength axis, and additionally over a region of $\pm 2$ pixels at each wavelength slice the applied atmospheric dispersion correction is updated. This results in data processing times of $\sim 10$ minutes and datacube volumes of $\sim 170$ Mb. The full covariance matrix is then recovered by interpolating between wavelength slices with covariance information, with the full covariance matrix produced in this way recovering $> 80$ per cent of the true covariance (Figure~\ref{covarhist}). The SAMI datacubes contain an additional \texttt{COVAR} extension which contains the covariance information sampled in this fashion. The \texttt{.fits} header of this extension records the wavelength slices (in pixels) at which the covariance was sampled, with the header item, \texttt{COVARLOCn} giving the wavelength slice of the $n^\mathrm{th}$ element along the wavelength axis of the sparsely sampled covariance matrix. An example implementation of the simple interpolation necessary to recover the full data is found in Appendix~\ref{pseudocode}.

The effectiveness of this covariance compression model is assessed by comparing the full covariance matrix to the modelled data on a spaxel by spaxel basis. Each spaxel is assessed across the 5$\times$5 covariance grid, normalised to unity centred on the spaxel (i.e., for unit variance). The difference between the true covariance matrix and the modelled data is then summed, both while retaining the sign of differences and also as an absolute sum of differences. A cumulative histogram of these sums is then generated (Figure~\ref{covarhist}). It is found that for 85\% of spaxels the missing covariance signal amounts to $< 10$\% the variance associated with each spaxel for the simple summation. When considering the summed absolute differences the missing covariance signal remains $< 20$\% of the variance for 80\% of spaxels. This level of accuracy is considered to be sufficient for most SAMI Galaxy Survey purposes. For specific individual cases in which higher accuracy is required, individual galaxy datacubes will be reformed with either finer sampling or full covariance array generation.

\section{Optimal dither strategy}
\label{optimaldither}
For a given fibre geometry within each SAMI IFU, and for a given number of dithered observations, it is necessary to determine the optimal dither pattern which provides the {\it most uniform coverage} of the output image map. In essence we wish to provide all parts of the combined output image with a uniform effective exposure time, i.e., we wish to have a uniform weight map.
The metric developed to optimise the alignment is to minimise the ratio of the standard deviation of flux weights in the output cube relative to the median flux weight.

The minimisation is restricted to a circular region which completely encompasses the {\it central} position of the dither pattern (typically the first observations).

Uniform coverage is trivially achieved by increasing the number of dither positions available. This is of course limited by finite observation time, CCD readout overheads, and the sensitivity requirement of achieving sky-limited dither exposures, all of which restricts the number of independent frames. Analysis reveals that while the optimal dither strategy lies within an extended plateau of parameter space (i.e., there are many essentially optimal strategies) it is surprisingly simple to select pathologically bad strategies that provide highly sub-optimal coverage and structure with a wide range of effective exposures times (or even image holes) across the final mosaic.

\begin{figure*}
\begin{center}
\includegraphics[width=150mm]{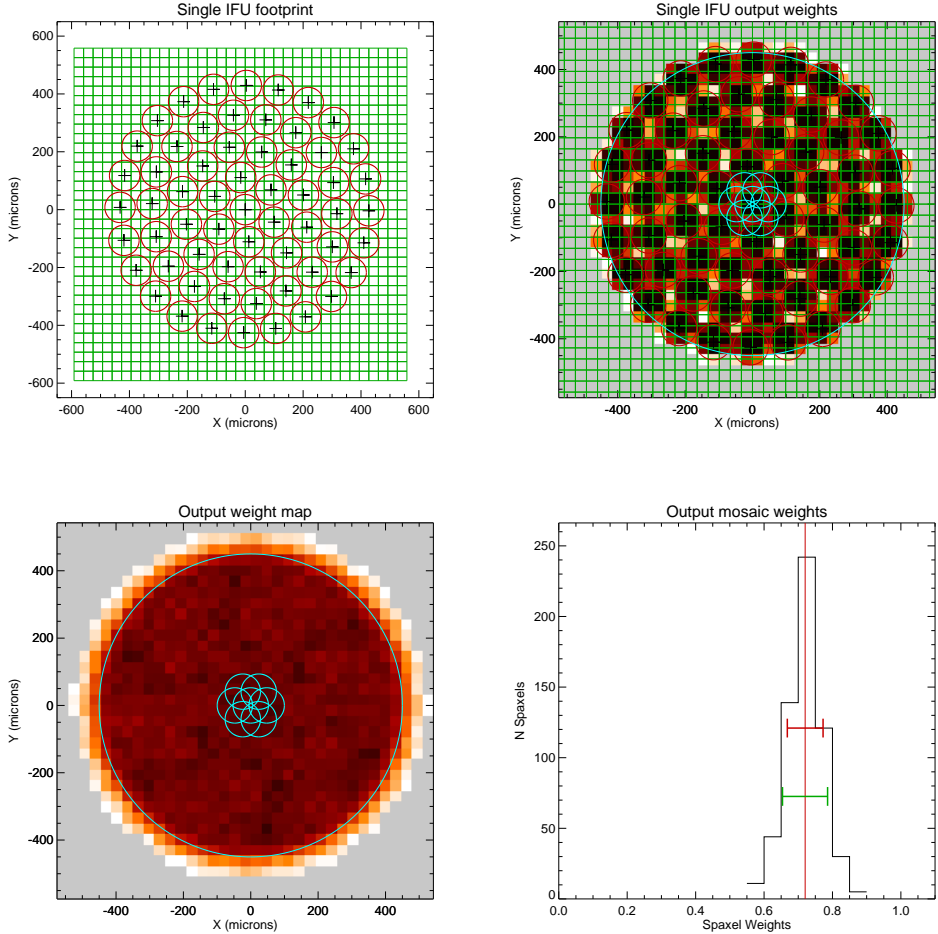}
\caption{\label{samifootprint} The SAMI data format and dither strategy.
Top-left) A single hexabundle is shown overlaid on a 0\farcs5 regular Cartesian grid. Each small circle represents an individual fibre within the IFU bundle.
Top-right) A drizzle remapping showing the relative weight of each output spaxel for a single observation. The hexagonal dither pattern for a single fibre core is also shown.
Bottom-left) The relative weight map for a 7-point hexagonal dither. The footprint of a single fibre in the dither pattern is shown overlaid.
Bottom-right) the histogram of relative weights for the simulated observation. The standard deviation (red) and interquartile range (green) are indicated.}
\end{center}
\end{figure*}

For the fibre arrangement of the SAMI hexabundles, a  seven-point hexagonal close-packed strategy, with a radial dither offset of 45\% of the fibre core diameter (i.e., 0\farcs72 for the 1\farcs6 SAMI fibres) was selected as the default SAMI observation mode (Figure~\ref{samifootprint}).
The impact of the rotation angle for this pattern was explored, noting that each SAMI hexabundle has an accurately known but distinct {\it lattice} structure. The effect was found to be negligible due to the high fill-factor of the SAMI IFS, the optimisation being dominated by local coverage around each fibre-core. For a system with lower fill-factor, the orientation of the offset pattern relative to the fibre-core lattice would likely be significant.

The most unexpected outcome was the realisation that the dither pitch should be less than half the fibre-core diameter in order to prevent the appearance of significant coverage holes at the centre of each core position. A further complication is the imprint of the dither strategy on the resolution of stacked datacubes. This issue is addressed in the next section.

\section{Recovering resolution via reduced drizzle drop-size}
\label{dropsize}
The main motivation for the original drizzle algorithm \citep{fruchter02} was to provide a means of recovering resolution in under-sampled $HST$/WFPC2 images and avoid introducing a convolution with the selected output pixel size which would further degrade the under-sampled PSF. Using a number of sub-pixel dithered observations of the input image, \citet{fruchter02} demonstrate that not only can one avoid significant degradation of the input image, but also that a significant fraction of the intrinsic resolution of the imaging system can be recovered despite the sub-critical sampling of each individual component frame. This is achieved by reducing the effective size of each input spaxel while scaling the flux to account for the reduced area (Figure~\ref{drizzleexample}). This reduction in the drizzle {\it drop-size} results in a smaller footprint for each input spaxels on the output grid. Each additional frame is drizzled onto the output grid but shifted by a fraction of the input spaxel true size and, with careful selection of the drop-size reduction, the process has been shown to result in high-fidelity images which faithfully represent the input image profile while minimising image degradation due to pixelisation by the observing system.

Drizzle is now routinely applied to imaging data from numerous sources and the soundness of the resampling has been confirmed a number of times \citep[e.g., ][and references therein]{koekemoer11}. However, in such cases the transformation has always been for a continuously sampled image. Below we explore the effects of drizzling with a reduced drop-size on the 73\% fill factor and under-sampled SAMI data (1\farcs6 fibres sampling with typical seeing of $\sim$2\farcs0).

\subsection{Model data for drizzle resampling}
In order to test the impact of the SAMI sampling on recovery of an image PSF, a series of model observations were created. Noiseless stellar profiles, modelled in the first instance as a 2D symmetric Gaussian profile, were generated assuming a seven-point hexagonal close-packed dither strategy. The model data were generated as RSS frames and then mosaiced into datacubes using the SAMI drizzle code. The RSS spectra were generated via numerical integration of the 2D PSF across the face of a model SAMI hexabundle using measured core position properties.

\subsection{Recovered PSF width}
In all cases a seven-point hexagonal dither was considered, a choice driven largely by the need to reach a final integration time total in individually sky limited observations.
A number of different drizzle drop sizes are also explored for resampling the simulated RSS data to construct  the datacube.

It is apparent from Figure~\ref{drizdropsize} that
using the full drop size generates a broadened profile with a reduced peak intensity. The data are degraded by an amount consistent with the smoothing by the fibre diameter. On reducing the drop size from 1\farcs6 to 0\farcs8 resolution is recovered although not to the full input image resolution.  The image degradation, an inevitable consequence of the subcritical sampling of the input image by the fibre footprint, is dependant on the input image FWHM. For the median seeing expected in survey quality 
SAMI data, this image blur results in an increase of the image FWHM of $\sim$0\farcs2 in simulated data.
Reducing the drizzle drop size below 50 percent of the fibre size does result in a more compact image, but the incomplete sampling due to the limited fill factor of the resulting seven-point dither pattern introduces artificial structure to the image. 

\subsection{Survey data image quality}
A limited quantity of observational data is available with which to confirm the simulated observations. Each SAMI Galaxy Survey plate contains a single calibration star. These standard stars are analysed in the same manner as the simulated data, with the observational seeing assessed in each individual RSS frame via forward modelling of the input stars and the final output image seeing measured directly from each stars dither combined datacube. datacubes are generated with an output spaxel size of 0\farcs5, with a drizzle drop size of 0\farcs8, a 50\% reduction. At the time of writing 242 individual RSS frames were available from 36 individual SAMI Galaxy Survey fields. This includes observations over a wide range of observational conditions including some periods of poor seeing. Data taken in the poorest conditions will not comprise part of the final SAMI Galaxy Survey, but is included in this analysis for completeness.

\begin{figure}
\begin{center}
\includegraphics[width=90mm]{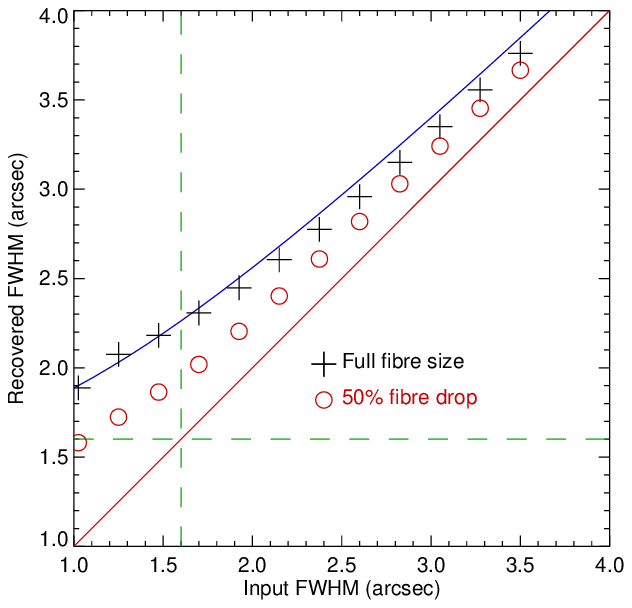}
\includegraphics[width=90mm]{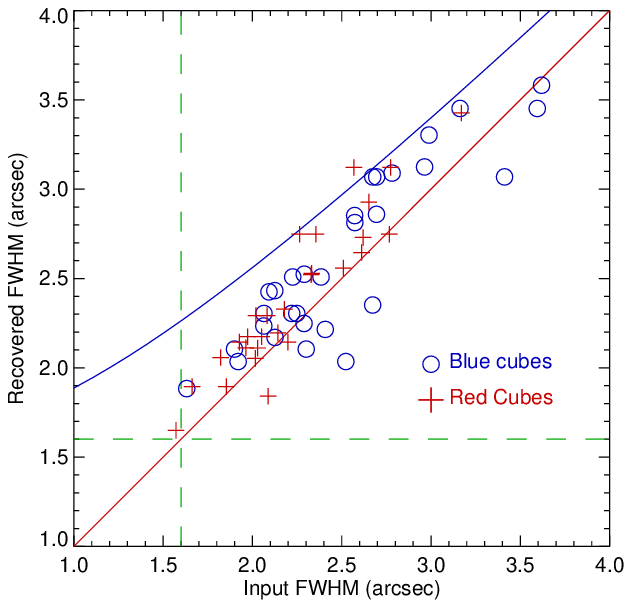}
\caption{\label{drizdropsize} Upper plot) The recovered resolution, with and without a reduction of the drizzle drop size is shown for simulated data. Simulated observations are generated with a range of Gaussian PSF FWHM. The full fibre diameter and a drizzle drop size of 0\farcs8 diameter (a 50\% reduction) are used to resample data onto an output spaxel size of 0\farcs5. A diagonal line marks the 1:1 correlation between input and output FWHM, while the curve corresponding to the quadrature sum of the input FWHM and the 1\farcs6 fibre diameter. Dashed vertical and horizontal lines mark this fibre diameter.
Lower plot) For the available standard stars taken from year one of the SAMI Galaxy Survey science observations, the median input seeing for each observing block (4-8 dithered frames) is compared with the seeing recovered from fitting to the final datacube of each star.}
\end{center}
\end{figure}

Figure~\ref{drizdropsize} compares the final recovered seeing for each star to the median of the input seeing for each data set. A chromatic term is visible between the red and blue datacubes, with the blue data typically 0\farcs2 poorer. This is consistent with the typical wavelength dependence of seeing over the two wavelength ranges. A minor AAT tracking error (identified through analysis of the year one SAMI data) is found to introduce a 10\% ellipticity in the year-1 SAMI data set. Remedial work is underway at the AAT to rectify the issues. FWHM measurements are taken as the geometric mean of a 2D elliptical Gaussian fit to the data. Fitting residuals are improved slightly, leading to a modest reduction in FWHM, if a 2D Moffat profile \citep{moffat69} is fitted but for consistency with the simulated test data the Gaussian form is presented. The measured cube FWHM values are found to lie above the 1:1 correlation, as expected. The departure from this correlation is not as marked as in the clean simulated data, largely due to the averaging effect for datasets with a marked seeing variation. In early SAMI data products no explicit resolution matching is performed beyond the simple rejection of markedly discrepant data sets.

The mean and standard deviation of input seeing values are found to be 2\farcs53$\pm$0\farcs60 in the blue and 2\farcs25$\pm$0\farcs43 in the red. The significant scatter is due to the inclusion of the full year-1 data set in this analysis.  For the resulting cubed data the corresponding values are  2\farcs61$\pm$0\farcs62 and  2\farcs40$\pm$0\farcs64. This analysis indicates a degradation of the input of not more than 0\farcs2 due to the alignment and cubing process when performed with the 50\% drizzle drop size reduction. In the context of the input fibre diameter of 1\farcs6 and typical AAT seeing, this is deemed satisfactory. As the SAMI Galaxy Survey progresses the median seeing of survey data will be improved via judicious flagging and rejection of poor seeing data.

\subsection{Flux scaling}
Reduction of the drizzle drop size requires a rescaling of the output cube values in order to preserve the flux calibration. Consider a linear reduction in the drop-size by a scale factor $\zeta$ (i.e., $\zeta$=0.5 for a drop size of 0\farcs8 scaled down from the SAMI fibre core size of 1\farcs6). We perform the drizzle resampling of the input flux RSS frame onto an output spaxel grid with the input fibre geometry set to incorporate the reduction factor. This produces the usual set of output data product, $C(r,\lambda), V(r,\lambda), $ \& $W(r,\lambda)$. Consider also the intermediate datacube prior to division by the weight map, $C'(r,\lambda)=C(r,\lambda) \times W(r,\lambda)$. If the total flux in the output cubes is calculated for a data set with and without the scale factor $\zeta$ clearly the values will differ by a factor of $\zeta^2$. This is a consequence of the smaller fibre foot prints covering fewer output spaxels and so less flux is distributed. In order to preserve the flux we find

\begin{align}
C'&=C/\zeta^2, \nonumber \\
V'&=V/\zeta^4, \nonumber \\
W'&=W/\zeta^2,
\end{align}

However, because the default SAMI data products are provided with the data and variance cubes divided by the weight map, in order to remove relative exposure time variations across the mosaic, the $\zeta^2$ scaling is effectively removed from the output datacube (the scaling being tracked in the weight array, $W$). This means that for most applications, users need not concern themselves with the weight maps.

\section{Conclusion}
To generate the data products necessary to realise the key science goals of the SAMI Galaxy Survey we have devised a scheme to generate 
Cartesian gridded datacubes from the irregularly sampled SAMI observational data products. Adapting established image processing techniques, we have demonstrated SAMI data can be accurately regularised, accounting for complications such as atmospheric refraction and dispersion, flux calibration and incomplete sampling of the focal plane. Image degradation due to the sparse sampling and reconstruction is limited to that inherent in the use of an input fibre core comparable in size to the natural image seeing scale. Photometric error information derived during data processing is accurately propagated as well as a compact representation of the complex covariance inevitable in the data. Using these techniques, the SAMI Galaxy Survey is exploring high-multiplex integral-field spectroscopy.

\section*{Acknowledgments}
The SAMI Galaxy Survey is based on observation made at the Anglo-Australian Telescope. The Sydney-AAO Multi-
object Integral-field spectrograph (SAMI) was developed jointly by the University of Sydney and the Australian Astronomical Observatory. The SAMI input catalogue is based on data taken from the Sloan Digital Sky Survey, the GAMA
Survey and the VST ATLAS Survey. The SAMI Galaxy Survey is funded by the Australian Research Council Centre of
Excellence for All-sky Astrophysics (CAASTRO, CE1100010200) and other participating institutions. The SAMI Galaxy Survey website is http://sami-survey.org. We thank the dedicated staff at the Anglo-Australian Telescope (AAT) whose support in interfacing SAMI with AAOmega is invaluable.

MSO \& JTA acknowledge the funding support from the Australian Research Council through a Super Science Fellowship (ARC FS110200023 \& FS110200013). SMC acknowledges the support of an ARC future fellowship (FT100100457).
ISK is the recipient of a John Stocker Postdoctoral Fellowship from the Science and Industry Endowment Fund (Australia).
LC acknowledges support under the Australian Research Councils Discovery Projects funding scheme (DP130100664).
CJW acknowledges support through the Marie Curie Career Integration Grant 303912.

This research made use of Astropy, a community-developed core Python package for Astronomy (Astropy Collaboration, 2013)\\

\footnotesize{

}

\clearpage

\appendix

\section{Code example for covariance recovery}
\label{pseudocode}
Presented below is a pseudo-code example of the covariance recovery process.
While the spectral datacube is not required in the reconstitution of the covariance matrix, four other components are: The spectral variance cube, the weight map, the compressed covariance cube, and, the \texttt{.FITS} header from the covariance cube extension. The procedures for regeneration of the covariance data is as follows:
\begin{itemize}
\item The covariance cube \texttt{.FITS} header is checked for the keyword \texttt{COVARMOD = Optimal}.
\item A full size array is created to store the reconstructed covariance.
\item The number of slices used to store the compressed covariance data is recovered from the covariance cube \texttt{.FITS} header via the \texttt{COVAR\_N} keyword.
\item The indices for the wavelength slices at which the covariance map has been stored in the compressed form are extracted from the covariance cube FITS header via the \texttt{COVARLOCn} keywords - where \texttt{n} is the index of the slice in the reduced array.
\item Running through each wavelength plane of the full size covariance array in turn, the array is populated with the corresponding slice from the compressed covariance array starting from element \texttt{0} of the compressed array, and moving to the next element of that array at the wavelength sliced indicated by \texttt{COVARLOCn}.
\item With the full size covariance array is now populated with the normalised data from the compressed array, one now loops over each wavelength slice in turn and multiplies the normalised mapping by the variance value, with the weight map normalisation removed, for each output spaxel.
\end{itemize}

\begin{table*}
\begin{verbatim}
# Python based pseudo-code to regenerate covariance array.
# Input are:
#                  The compressed covariance array
#                  Its FITS header
#                  The full variance array
#                  The full weight map 
# The full reconstructed covariance array is returned.

def reconstruct_covariance(var_array,covar_array_red,weight_array,covar_header):
    # Reconstruct the full covariance array from the reduced covariance
    # information stored in a standard cube
    
    if covar_header['COVARMOD'] != 'optimal':
        raise Exception('This cube does not contain covariance information in the optimal format')
    
    # Create an empty full covariance cube
    # This has dimensions of [Wavelength, Covariance scale, Covariance scale, Cube size, Cube size]
    covar_array_full = np.zeros([2048,5,5,50,50])
    
    # Populate the full covariance cube with covariance maps from reduced array
    n_covar = covar_header['COVAR_N']
    for i in range(n_covar):
        slice = covar_header['COVARLOC'+str(i+1)]
        covar_array_full[slice,:,:,:,:] = covar_array_red[i,:,:,:,:]
    
     # Fill values between calculated covariance map slices with
     # the last calculated value
    lowest_point = np.min(np.where((covar_array_full[1:,2,2,25,25] != 0.0) &
                                   (np.isfinite(covar_array_full[1:,2,2,25,25])))[0]) + 1
    for i in range(2048):
        if np.sum(np.abs(covar_array_full[i,:,:,:,:])) == 0:
            if i < lowest_point:
                covar_array_full[i,:,:,:,:] = covar_array_full[lowest_point,:,:,:,:]
            else:
                covar_array_full[i,:,:,:,:] = covar_array_full[i-1,:,:,:,:]
                   
    # For each wavelength slice at each spaxel,
    # scale the normalised covariance maps by the appropriate variance removing the
    # after removing the relative exposure time weight map normalisation.
    for i in range(2048):
        for x in range(50):
            for y in range(50):
        covar_array_full[i,:,:,x,y] = covar_array_full[i,:,:,x,y] * var_array[i, x,y] * (weight_array[i,x,y]**2)
                                       
    return covar_array_full
\end{verbatim}
\end{table*}

\bsp

\label{lastpage}


\begin{thebibliography}{99}
\bibitem[Abazajian et al.(2009)]{abazajian09}
Abazajian, K.~N., Adelman-McCarthy, J.~K., Ag\"ueros, M.~A., Allam, S.~S.; Allende P.~C., An, D., Anderson, K.~S.~J., Anderson, S.~F., et al., 2009, ApJS, 182, 543
\bibitem[Allen et al.(2014a)]{allen14a}
Allen, J.~T., et al. 2014a, Astrophysics Source Code Library, ascl:1407.006
\bibitem[Allen et al.(2014b)]{allen14b}
Allen, J.~T., Croom, S.~M., Konstantopoulos, I.~S., Bryant, J.~J., Sharp, R.;, Cecil, G.~N., Fogarty, L.~M.~R., Foster, C., et al., 2014, arXiv:astro-ph1407.6068A	
\bibitem[Bacon et al.(2001)]{bacon01}
Bacon, R., Copin, Y., Monnet, G., Miller, B.~W., Allington-Smith, J.~R., Bureau, M., Carollo, C.~M., Davies, R.~L., et\,al., 2001, MNRAS, 326, 23
\bibitem[Bland \& Tully(1989)]{bland89}
Bland, J., \& Tully, R.~.B, 1989, AJ, 98, 723
\bibitem[Bland-Hawthorn (2011)]{bland-hawthorn11}
Bland-Hawthorn, J., Bryant, J., Robertson, G., Gillingham, P., O'Byrne, J., Cecil, G., Haynes, R., Croom, S., et al., 2011, O.~Expr, 19, 2649
\bibitem[Blanton \& Moustakas(2009)]{blanton09}
Blanton, M.~R., Moustakas, J., 2009, ARA\&A, 47, 159
\bibitem[Brough et al.(2013)]{brough13}
Brough, S., Croom, S., Sharp, R., Hopkins, A.~M., Taylor, E.N., Baldry, I.~K., Gunawardhana, M.~L.~P., Liske, J., 2013, et al., MNRAS, 435, 2903
\bibitem[Bryant et al.(2014a)]{bryant14a}
Bryant, J.~J., Bland-Hawthorn, J., Fogarty, L.~M.~R., Lawrence, J.~S., Croom, S.~M., 2014a, MNRAS, 438, 869
\bibitem[Bryant et al.(2014b)]{bryant14b}
Bryant, J.~J., Owers, M.~S., Robotham, A.~S.~G., Croom, S.~M., Driver, S.~P., Drinkwater, M.~J., Lorente, N.~P.~F., Cortese, L., et al., 2014b, arXiv:astro-ph1407.7335B	
\bibitem[Bryant (2012)]{bryant12}
Bryant, J.~J., Bland-Hawthorn, J., Lawrence, J., Croom, S. Fogarty, L.~M., Goodwin, M., Richards, S., Farrell, T., et al., 2012, SPIE 8446E, 0XB
\bibitem[Bryant et al.(2011)]{bryant11}
Bryant, J.~J., O'Byrne, J.~W., Bland-Hawthorn, J., Leon-Saval, S.~G. 2011, MNRAS, 415, 2173		
\bibitem[Castelli \& Kurucz(2004)]{CastelliKurucz2004}
Castelli, F., Kurucz, R.~L., 2004, arXiv:astro-ph/0405087
\bibitem[Cappellari et\,al.(2011)]{cappellari11}
Cappellari, M., Emsellem, E., Krajnovi\'c, D., McDermid, R.~M., Scott, N., Verdoes, K.~G.~A., Young, L.~M., Alatalo, K., et\,al., 2011, MNRAS, 413, 813
\bibitem[Colless et al.(2001)]{colless01}
Colless,~M., Dalton,~G., Maddox,~S. Sutherland,~W., Norberg,~P., Cole,~S., Bland-Hawthorn,~J., Bridges,~T., et\,al., 2001, MNRAS, 328, 1039
\bibitem[Croom et al.(2012)]{croom12}
Croom, S.~M., Lawrence, J.S., Bland-Hawthorn, J., Bryant, J.~J., Fogarty, L., Richards, S., Goodwin, M., Farrell, T., et al., 2012, MNRAS, 421, 872
\bibitem[Drinkwater et al.(2012)]{drinkwater12}
Drinkwater,~M.~J., Jurek,~R.~J., Blake,~C., Woods,~D., Pimbblet,~K.~A., Glazebrook,~K., Sharp,~R, Pracy,~M.~B., 2010, MNRAS, 401, 1429
\bibitem[Driver et al.(2006)]{driver06}
Driver, S.~P., Allen. P.~D., Graham, A.~W., Cameron, E., Liske, J., Ellis, S.~C., Cross, N.~J.~G., De Propris, R., et al., 2006, MNRAS, 368, 414 
\bibitem[Driver et al.(2011)]{driver11}
Driver, S.~P., Hill, D.~T., Kelvin, L.~S., Robotham, A.~S.~G., Liske, J. Norberg, P.,, Baldry, I.~K., Bamford, S.~P, et al., 2011, MNRAS, 413, 971
\bibitem[Eisenstein et al.(2005)]{eisenstein05}
Eisenstein, D.~J., Zehavi, I., Hogg, D.~W., Scoccimarro, R., Blanton, M.~R., Nichol, R.~C., Scranton, R., Seo, H.~-J., et\,al., 2005, ApJ, 633, 560
\bibitem[Filippenko(1982)]{filippenko82}
Filippenko, A.~V. 1982, PASP, 94, 715	
\bibitem[Fogarty et al.(2012)]{fogarty12}
Fogarty, L.~M.~R., Bland-Hawthorn, J., Croom, S.~M., Green, A.~W., Bryant, J.~J., Lawrence, Jon S., et al. 2012, ApJ, 761, 169
\bibitem[Fogarty et al.(2014)]{fogarty14}
Fogarty, L.~M.~R., Scott., N., Owers, M.~S., Brough, S., Croom, S.~M., Pracy, M.~B., Houghton, R.~C.~W., Bland-Hawthorn, J., et al., 2014, MNRAS, 443, 485
\bibitem[Fruchter \& Hook(2002)]{fruchter02}
Fruchter, A.S., Hook, R.N., 2002, PASP, 114, 144
\bibitem[Husemann et al.(2013)]{husemann13}
Husemann, B., Jahnke, K., S\'anchez, S.~F., Barrado, D., Bekeraite, S., Bomans, D.~J., Castillo-Morales, A., Catal\'an-Torrecilla, C., et al.  2013, A\&A, 549A, 87
\bibitem[Ho et al.(2014)]{ho14}
Ho, I.-T., et al., 2014, arXiv:astro-ph1407.2411
\bibitem[Hopkins et al.(2013)]{hopkins13}
Hopkins, A.~M., Driver, S.~P., Brough, S., Owers, M.~S., Bauer, A.~E., Gunawardhana, M.~ L.~P., Cluver, M.~E., Colless, M., et al.,  2013, MNRAS, 430, 2047
\bibitem[Huchra et al.(1999)]{huchra99}
Huchra, J.~P., Vogeley, M.~S., Geller, M.~J., 1999, ApJS, 121, 287
\bibitem[Husemann, et al.(2012)]{husemann12}
Husemann, B., Kamann, S., Sandin, C., S\'anchez, S.~F., Garc\'ia-Benito, R., Mast, D., 2012, A\&A, 545, 137
\bibitem[Jones et al.(2009)]{jones09}
Jones, D.~H.,,Read, M~.A.,Saunders, W., Colless, M., Jarrett, T., Parker, Q.~A., Fairall, A.P., Mauch, T., et\,al.,  2009, MNRAS, 399, 683
\bibitem[Kelson(2003)]{kelson03}
Kelson, D.D., 2003, PASP, 115, 688
\bibitem[Koekemoer et al.(2011)]{koekemoer11}
Koekemoer, A.~M., Faber, S.~M., Ferguson, H.~C., Grogin, N.~A., Kocevski, D.~D., Koo, D.C., Lai, K., Lotz, J.~M., et al., 2011, ApJS, 197, 36
\bibitem[Lawrence et al.,(2012)]{lawrence12}
Lawrence, J., Bland-Hawthorn, J., Bryant, J., Brzeski, J., Colless, M., Croom, S. Gers, L. Gilbert, J., et al., 2012, SPIE, 8446E, 53 
\bibitem[Markwardt(2009)]{markwardt09}
Markwardt, C.B., 2009, ASPC, 411, 251 (arXiv:0902.2850v1)
\bibitem[Roth et al.(2005)]{roth05}
Roth, M~.M., Kelz, A., Fechner, T., Hahn, T., Bauer, S.~-M., Becker, T., B\"ohm, P., Christensen, L., et\,al., 2005, PASP, 117, 620
\bibitem[S\'anchez et al.(2012)]{sanchez12}
S\'anchez, S.~F., Kennicutt, R.~C., Gil de Paz, A., van de Ven, G., V\'ilchez, J.~M., Wisotzki, L., Walcher, C.~J.,; Mast, D., et\.al., 2012, A\&A, 538A, 8
\bibitem[Moffat(1969)]{moffat69}
Moffat, A.~F.~J., 1969, A\&A, 3, 455
\bibitem[Rich et al.(2012)]{rich12}
Rich, J.~A., Torrey, P., Kewley, L.~J., Dopita, M.~A., Rupke, D.~S.~N.,  2012, ApJ, 753, 5
\bibitem[Richards et al.,(2014)]{richards14}
Richards, S.~N., et al., in prep.
\bibitem[Saunders et al.,(2004)]{saunders04}
Saunders, W., Bridges, T., Gillingham, P., Haynes, R., Smith, G.~A., Whittard, J.~D., Churilov, V., Lankshear, A., et al.,  2004, SPIE, 5492, 389
\bibitem[Sharp \& Birchall(2010)]{sharp10a}
Sharp, R., Birchall, M.~N., 2010, PASA, 27, 91
\bibitem[Sharp \& Bland-Hawthorn(2010)]{sharp10c}
Sharp, R.~G., \& Bland-Hawthorn, J., 2010, ApJ, 711, 818
\bibitem[Sharp, Brough \& Cannon(2013)]{sharp13}
Sharp, R., Brough, S., Cannon, R.~D., 2013, MNRAS, 428, 447
\bibitem[Sharp \& Parkinson(2010)]{sharp10b}
Sharp, R., Parkinson, H., 2010, MNRAS, 408, 2495	
\bibitem[Sharp, Saunders, Smith, et al.,(2006)]{sharp06}
Sharp, R., Saunders, W., Smith, G., Churilov, V., Correll, D., Dawson, J., Farrel, T., Frost, G. et al., 2006, SPIE 6269E, 14
\bibitem[van Dokkum(2001)]{vandokkum01}
van Dokkum, P.~G., 2001, PASP, 113, 1420
\bibitem[Veilleux et al.(2003)]{veilleux03}
Veilleux, S., Shopbell, P.~L., Rupke, D.~S., Bland-Hawthorn, J., Cecil, G., 2003, AJ, 126, 2185
\bibitem[Welikala et al.(2008)]{welikala08}
Welikala, N., Connolly, A.~J., Hopkins, A.~M., Scranton, R., Conti, A., 2008, ApJ, 677, 970
\bibitem[York et al.(2000)]{york00}
York, D.~G., Adelman, J., Anderson, J.~E.,~Jr., Anderson, S.~F., Annis, J., Bahcall, N.~A., Bakken, J.~A., Barkhouser, R.\ et al.,2014, AJ, 443, 485
\end{thebibliography}
\end{document}